\documentclass[10pt]{elsarticle}

\usepackage[T1]{fontenc}
\usepackage{times}%
\PassOptionsToPackage{hyphens}{url}\usepackage[hidelinks]{hyperref}
\usepackage{graphicx}
\usepackage{epstopdf}
\usepackage{color}
\usepackage{listings}
\usepackage{amssymb}
\usepackage{float}
\usepackage{array}
\usepackage{calc}
\usepackage{threeparttablex}
\hypersetup{
  colorlinks   = true, 
}

\usepackage{booktabs,adjustbox}
\usepackage{soul} 
\usepackage[normalem]{ulem}
\usepackage{placeins}
\usepackage{caption}
\usepackage[font=scriptsize]{subcaption}
\usepackage{multirow}
\usepackage{ragged2e}
\usepackage{array}
\usepackage[utf8]{inputenc}
\usepackage[T1]{fontenc}

\usepackage{forest}
\usepackage{epsfig}
\usepackage{verbatim}
\usepackage{lipsum}

\usepackage{algorithm}
\usepackage{algorithm,lipsum,etoolbox}
\usepackage[noend]{algpseudocode}
\usepackage{adjustbox}
\usepackage{newfloat}
\usepackage{etoolbox}
\usepackage{graphicx}
\usepackage{xfrac}
\usepackage{tikz}
\usepgflibrary{arrows}
\newcolumntype{P}[1]{>{\RaggedRight\hspace{0pt}}p{#1}}
\usepackage{amsmath}
\usepackage{longtable}

\usepackage{enumitem}
\usepackage{color,amsmath,amssymb}

\makeatletter
\expandafter\patchcmd\csname\string\algorithmic\endcsname%
      {\labelwidth 0.5em}{\labelwidth0pt\labelsep0pt}{}{}

\makeatletter
\newif\if@algorithm
\patchcmd{\@float@Hx}
  {\@nodocument}
  {\@nodocument%
   \ifnum\pdfstrcmp{#1}{algorithm}=0 
     \@algorithmtrue
     \setlength{\columnwidth}{\algorithmwidth}
   \fi}
  {}{}
\renewcommand\float@makebox[1]{%
  \hbox{%
    \if@algorithm\hspace*{-.5\dimexpr\algorithmwidth-\textwidth}\fi%
    \vbox{\hsize=#1 \@parboxrestore
      \@fs@pre\@fs@iftopcapt
        \ifvoid\@floatcapt\else\unvbox\@floatcapt\par\@fs@mid\fi
        \unvbox\@currbox
      \else\unvbox\@currbox
        \ifvoid\@floatcapt\else\par\@fs@mid\unvbox\@floatcapt\fi
      \fi\par\@fs@post\vskip\z@}}}

\makeatother

\newlength{\algorithmwidth}
\AtBeginDocument{\setlength{\algorithmwidth}{1.2\textwidth}}
\usepackage{amsthm}

\usepackage [english]{babel}
\usepackage [autostyle, english = american]{csquotes}
\MakeOuterQuote{"}

\journal{Computers \& Security}

\date{}
\begin{document}

\begin{frontmatter}

\title{Dexteroid: Detecting Malicious Behaviors in Android Apps Using Reverse-Engineered Life Cycle Models}

\maketitle
Mohsin Junaid, Donggang Liu, David Kung Dexteroid: Detecting malicious behaviors in Android apps using reverse-engineered life cycle models, Computers \& Security, Volume 59,Pages 92-117, ISSN 0167-4048 \url{http://dx.doi.org/10.1016/j.cose.2016.01.008}, June 2016.

 \author{Mohsin Junaid\corref{cor1}}\ead{mohsin.junaid@mavs.uta.edu}
  \author{Donggang Liu\corref{cor3}}\ead{dliu@uta.edu}
   \author{David Kung\corref{cor2}}\ead{kung@uta.edu}
 \address{Department of Computer Science and Engineering, University of Texas at Arlington, Arlington, United States}

\cortext[cor1] {Corresponding Author}

\begin{abstract}

The amount of Android malware has increased greatly during the last few years. Static analysis is widely used in detecting such malware by analyzing the code without execution. The effectiveness of current tools relies on the app model as well as the malware detection algorithm which analyzes the app model. If the model and/or the algorithm is inadequate, then sophisticated attacks that are triggered by specific sequences of events will not be detected.

This paper presents a static analysis framework called Dexteroid, which uses reverse-engineered life cycle models to accurately capture the behaviors of Android components. Dexteroid systematically derives event sequences from the models, and uses them to detect attacks launched by specific ordering of events. A prototype implementation of Dexteroid detects two types of attacks: (1) leakage of private information, and (2) sending SMS to premium-rate numbers. A series of experiments are conducted on 1526 Google Play apps, 1259 Genome Malware apps, and a suite of benchmark apps called DroidBench and the results are compared with a state-of-the-art static analysis tool called FlowDroid. The evaluation results show that the proposed framework is effective and efficient in terms of precision, recall, and execution time.

\end{abstract}


\end{frontmatter}

\section{Introduction}

Android is the most popular mobile OS, with nearly 80\% of the market share \cite{IDCReport}. This attracts malware attacks such as leakage of sensitive information and sending SMS to premium-rate numbers \cite{FSecureReport, APFeltSurvey}. These attacks can be implemented by well-crafted malicious apps, or advertisement libraries used by Android apps \cite{adLibraries}. To combat such attacks, researchers have developed static analysis, dynamic analysis, and permission-based techniques. Static analysis detects malicious behaviors by analyzing the app code without execution \cite{LeakMiner, FlowDroid, AppoScopy, myDroid, Chex}. The aim is to represent program logic in some models (such as control flow graphs) and analyze such models to detect possible attacks. Dynamic analyses typically force the program to execute a set of carefully selected paths and analyze the results to detect malware attacks \cite{TaintDroid, VetDroid, DroidScope, DroidPF}. In addition to these, Android provides a permission-based mechanism \cite{permissions}, which requires user approval to access resources such as deviceID and internet. 

This paper focuses on static analysis. It is motivated by recent success of \cite{LeakMiner, FlowDroid, DroidSafe, AppoScopy, Chex, myDroid} in detecting Android malware. All of these techniques perform analysis based on the Android-supplied life cycle models, which can be represented by a set of state machines. The state machines model the state-dependent behaviors of Android components, that is, Activity, Service, BroadcastReceiver, and ContentProvider. A state represents the status of a component; a transition represents a callback invoked by Android. In particular, LeakMiner \cite {LeakMiner} and FlowDroid \cite{FlowDroid} derive a control flow graph (CFG) based on life cycle model. The nodes of the CFG represent callbacks in the life cycle model and the edges define the order that the callbacks can be invoked. Taint analysis is then performed on paths of the CFG to detect information leakage. Unlike LeakMiner and FlowDroid, DroidSafe \cite{DroidSafe} performs flow-insensitive taint analysis and analyzes all possible orderings of callbacks in the life cycle model to detect information-flows in Android apps. 

An Android-supplied life cycle model provides a high-level abstraction of the Android component behavior. However, this abstraction is also a limitation because it does not capture all states and transitions implemented in Android. Furthermore, the life cycle model does not specify guard conditions that govern the invocation of a callback. Thus, an analysis method based on an Android-supplied life cycle model may not detect attacks that exploit such omissions (the motivating example presented in the next section illustrates this). To fill this gap, we perform reverse-engineering to reconstruct the life cycle models that capture the omitted states and transitions, as well as the guard conditions. We then propose the Dexteroid framework. This framework systematically derives event sequences from the reverse-engineered life cycle models, obtains callback sequences from the event sequences, and performs taint analysis on the callback sequences to detect malicious behaviors.

The contributions of this paper are summarized as follows. First, Dexteroid detects attacks that are missed by existing tools, due to the inclusion of omitted states and transitions in the life cycle models. Second, to the best of our knowledge, this research is the first that uses event sequences (derived from a life cycle model) to generate callback sequences for analysis. Although callback sequences are considered in previous studies, most of such sequences will not occur in real apps because they do not satisfy the guard conditions. As a consequence, this could lead to a high false positive rate. Third, our research is also the first to systematically generate and analyze permutations of event sequences. This is because certain attacks can only be detected by analyzing more than one event sequences in a given order. Finally, we have designed and implemented a prototype of Dexteroid and conducted extensive experiments. The results show that the proposed framework significantly improves both precision and recall.

The rest of the paper is organized as follows. Section 2 provides background on Android components, and a motivating example to show the limitation of an Android-supplied life cycle model of a component. Section 3 presents Dexteroid framework. Section 4 provides the implementation details of Dexteroid and Section 5 presents experimental evaluation. Section 6 describes the related work. The last section concludes the paper and discusses future work. 

\section {Background and Motivating Example}

Android defines four basic components to develop Android applications: (1) \emph{Activity} implements the application logic as well as provides an interactive screen to the user; (2) \emph{Service} performs long-running operations and runs in the background; (3) \emph{BroadcastReceiver} responds to system-wide notifications; and (4) \emph{ContentProvider} manages structured data. 

Each component has a life cycle. That is, it goes through a series of phases, possibly iterates some of the phases, during its life time. For example, an activity goes through three phases when it becomes visible, partially visible and completely hidden to the user. The transitions of a component from one phase to another are caused by the occurrences of events, which trigger the invocations of callback methods or callbacks in short. In the following sections, we define important concepts shown in italic font in the sentences that define them. These concepts will be used throughout the paper.

\subsection{Basic Definitions}

An \emph{event} is some happening of interest; it can be an external event, or an internal event. An \emph{external event} occurs when the app user submits a request, or performs an action, such as tapping a button on an activity's user interface. The device hardware captures such events and delivers them to the operating system. An \emph{internal event} is generated by the operating system when certain condition, such as low on memory, becomes true. Such events are captured by the operating system. Whenever an external or internal event occurs, the operating system calls functions of the components. These are \emph{callback functions} or \emph{callback methods}. More specifically, a \emph{callback method} is a method of a component that is invoked by the operating system in response to an internal or external event. Table \ref{tabEventsCallbacks} shows example user operations, the corresponding events, and callback sequences that are invoked by the OS in an activity.

\begin{table}[ht] 
\caption{Sample User Operations, Triggered Events, and Invoked Callbacks in An Activity}
\centering 
\scriptsize
\begin{tabular}{l l l} 
\hline
 Operation performed & Triggered event & Invoked callback sequences\\  
\hline 
(1) Tap on app icon to start an activity & createActivity & {\ttfamily onCreate()}, {\ttfamily onStart()}, {\ttfamily onResume()} \\
(2) After (1), press Back button on the device & backPress & {\ttfamily onPause()}, {\ttfamily onStop()}, {\ttfamily onDestroy()} \\
(3) After (1), press Home button on the device & overlapActivity & {\ttfamily onPause()}, {\ttfamily onStop()} \\
(4) After (3), start the app from launcher menu & restartActivity & {\ttfamily onRestart()}, {\ttfamily onStart()}, {\ttfamily onResume()} \\

\hline 
\end{tabular} 
\label{tabEventsCallbacks} 
\end{table}

\subsection{Android-Supplied Activity Life Cycle Model}

\begin{figure}[ht!]
\centering
\includegraphics[width=12cm]{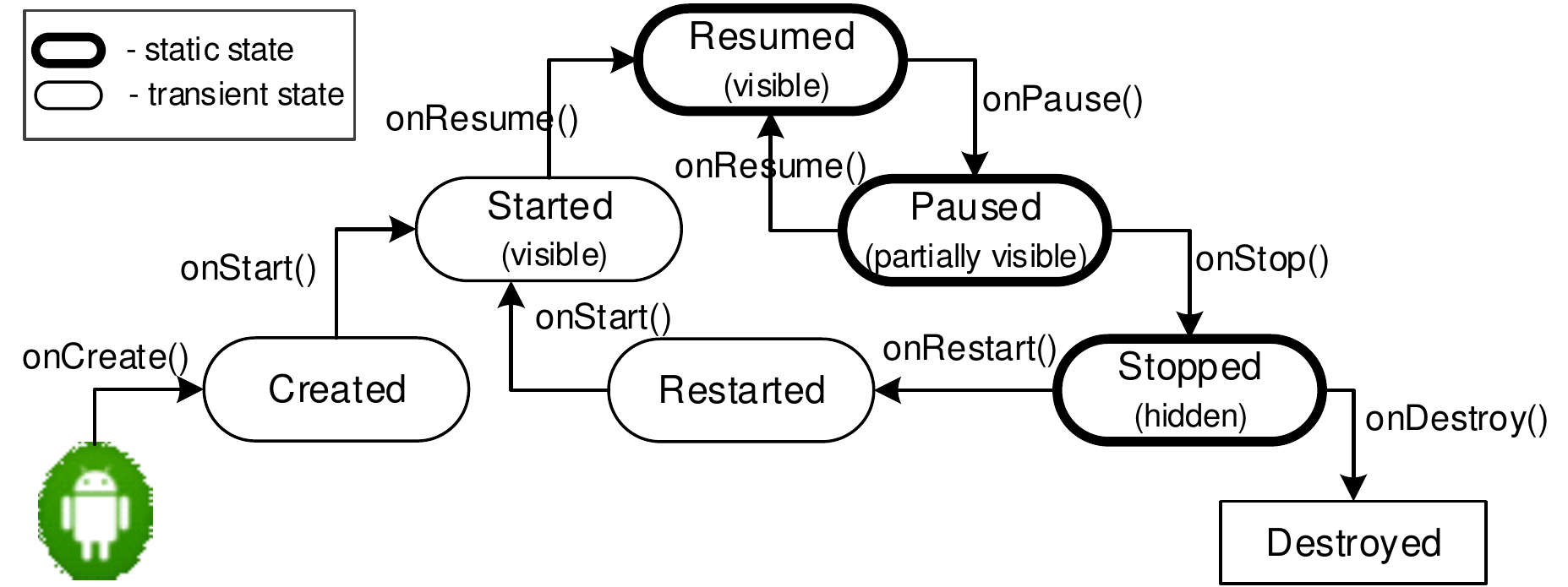}
 \caption{Android-Supplied Activity Life Cycle Model \cite{activityLifeCycle}}
  \label{coreLifeCycle}
\end{figure}

The life cycle of a component can be modeled by a state machine. The phases of the component are represented by states, and transitions between phases are represented by state transitions. Figure \ref{coreLifeCycle} shows the Android-supplied activity life cycle model \cite{activityLifeCycle} which describes the state-dependent behavior of an Android activity. It contains eight states including one initial state (Android robot icon) and nine state transitions. Each state transition in the model is caused by the invocation of a callback by the operating system. These callbacks are called \emph{life cycle callbacks}. For example, when a user opens or starts a new activity, createActivity event is triggered as shown in first operation of Table \ref{tabEventsCallbacks}. The system invokes the callback sequence (e.g., {\ttfamily onCreate()}, {\ttfamily onStart()} and {\ttfamily onResume()}) in the activity and in response, the activity makes transitions to Created, Started and Resumed state, respectively. Activity waits for user interaction in the Resumed state. A user can trigger many events in this state which can make activity visit different states. 

Activity states are classified into static states and transient states \cite{activityLifeCycle}. A \emph{static state} requires an external event to cause the invocation of a callback method which in turn triggers a state transition. Thus, an activity can stay in a static state indefinitely, depending upon when the external event arrives. The state transition for a \emph{transient state} takes place automatically and immediately after its associated callback method is invoked. For example, when the activity visits Created state, it automatically moves to Started state (both are transient states) and then moves to Resumed state (a static state). Figure \ref{coreLifeCycle} shows both static states (with bold border lines) and transient states. According to this model, the user can interact with the activity only when it is visible to the user in Resumed state.

There are seven life cycle callbacks in activity life cycle model shown in Figure \ref{coreLifeCycle}. In addition to these callbacks, other types of methods which are defined in Android applications are activity user-interface (AUI) callbacks, miscellaneous callbacks and developer-defined methods. \emph{Activity user-interface callbacks} or \emph{AUI callbacks} are invoked by the operating system in response to external events on AUI elements. For example, a typical user interaction of clicking a button triggers an event which causes the system to invoke its registered callback (e.g., {\ttfamily onButtonClicked()}). Since only an activity can provide user-interface for an application, AUI callbacks are defined in the activity code. However, they can be registered with a layout xml file or dynamically within the activity code. \emph{Miscellaneous callbacks} can be invoked by internal or external events and are not part of life cycle model of a component. For example, a user tap on the screen causes the system to invoke {\ttfamily onUserInteraction()} callback. Similarly, the system can trigger a `lowMemory' event to invoke {\ttfamily onLowMemory()} callback in both activity and service components. A \emph{developer-defined method} is a method which is executed by the system only when there is an explicit call to such a method. Such methods can be invoked from callback methods or other developer-defined methods.

The life cycle models of BroadcastReceiver and ContentProvider components are straightforward: each contains one state and a callback method ({\ttfamily onReceive()} and {\ttfamily onCreate()}, respectively). Activity is prevalent in Android apps and handles all types of (above-defined) methods which makes its life cycle model more complex compared to other components. Thus, we first discuss activity component and then service component in detail in this paper.

\subsection {Motivating Example}

\begin{figure}[ht]
\centering
\begin{subfigure}{.5\textwidth}
  \centering
  \includegraphics[width=30mm, height=37mm]{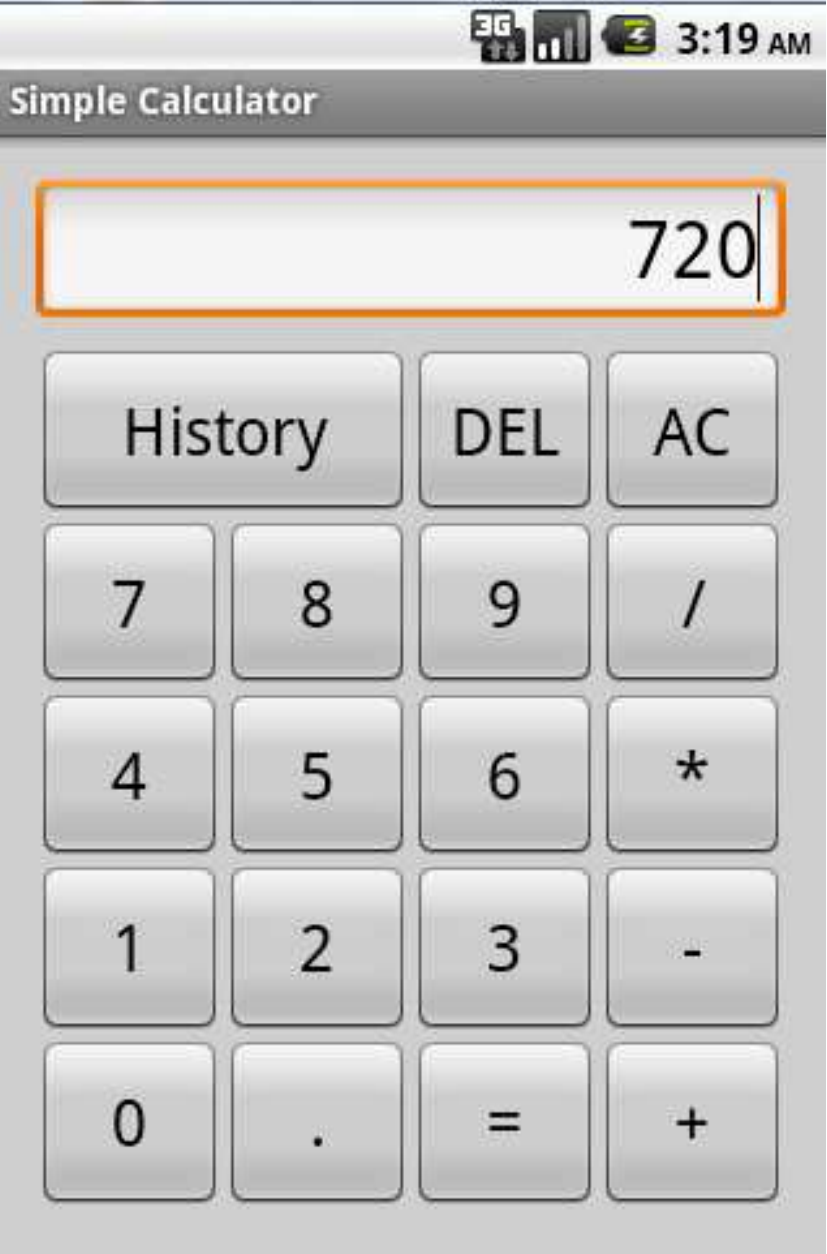}
  \scriptsize
  \caption{A user can perform many calculations here}
  \label{fig:calc}
\end{subfigure}%
\begin{subfigure}{.5\textwidth}
  \centering
  \includegraphics[width=30mm, height=37mm]{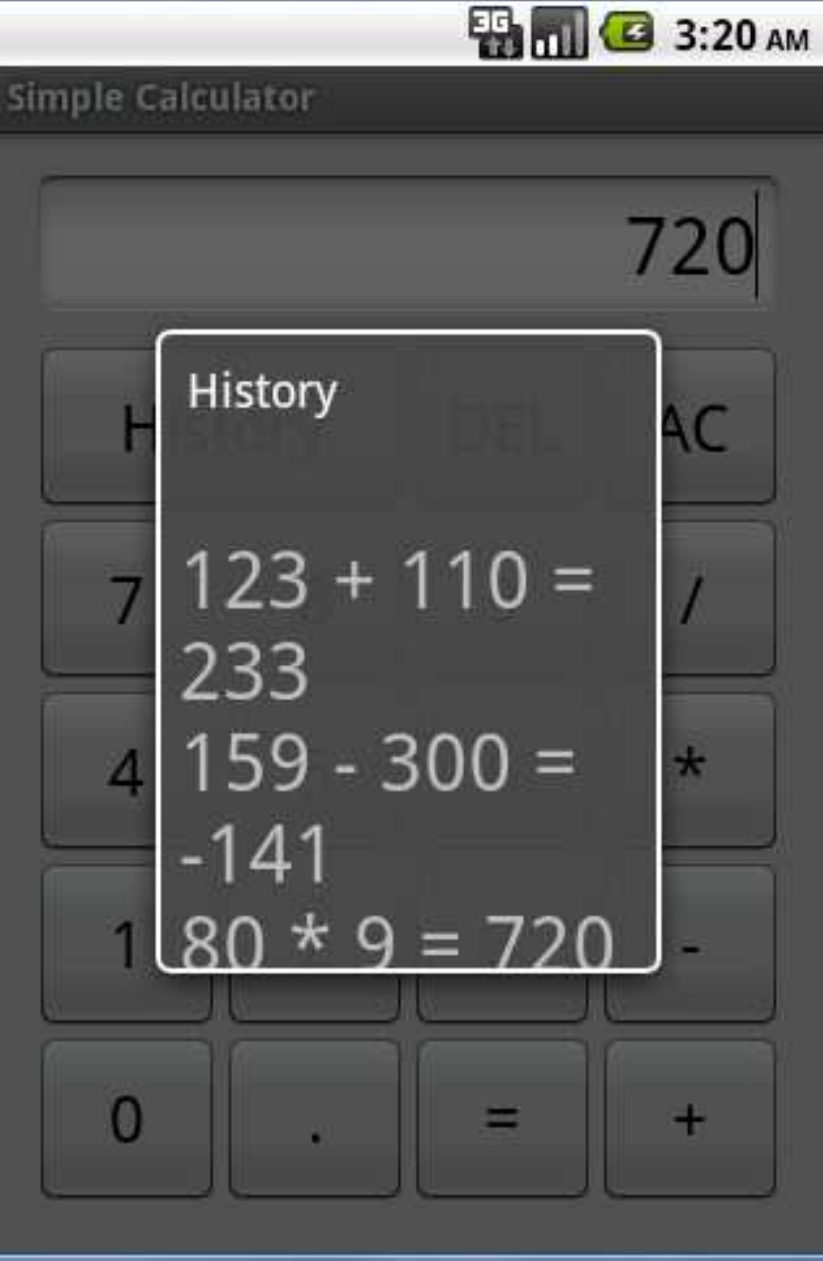}
  \caption{Calculations history is shown in a dialog activity}
   \label{fig:hist}
\end{subfigure}
\caption{A Calculator Application}
\label{motivatingExample}
\end{figure}

An attacker can launch a targeted attack using program flows overlooked by the analysis tools. The motivating example shows that an attack can be designed to launch by invoking a set of callbacks in a specific order. It further shows that Android-supplied activity life cycle model can be exploited to launch such attacks due to its omission of certain states and transitions. The example is a calculator app shown in Figure \ref{fig:calc}, in which a user can perform different operations, and then view their history in a dialog activity as shown in Figure \ref{fig:hist}. Listing \ref{listing1} shows the app code. To save space, non-essential code such as calls to super class functions and exception handling is omitted.

Consider the following sequence of operations performed by the user: (1) User starts an activity for the first time. (2) The user performs a few calculations and taps History button. History dialog activity is displayed as shown in Figure \ref{fig:hist}. The user taps Back button on the device to get back to {\ttfamily MainActivity}. (3) The user presses Home button on the device and later opens the app again from launcher menu. (4) The user rotates the device from portrait view to landscape view for better view of the app layout. 

Each of the above user operations triggers a sequence of one or more events which causes the system to invoke a sequence of callback methods in the activity. The respective callback sequences for these operations are: (1) {\ttfamily onCreate()} and {\ttfamily onResume()}. (2) {\ttfamily onBtnClicked()}, {\ttfamily onUserLeaveHint()}, {\ttfamily onSaveInstanceState()}. This callback sequence obtains deviceID (line 18 of Listing \ref{listing1}) and stores it in {\ttfamily d1} variable. Upon returning to the activity, {\ttfamily onResume()} is invoked by the operating system. (3) {\ttfamily onUserLeaveHint()}, {\ttfamily onSaveInstanceState()}, and {\ttfamily onResume()}. This callback sequence copies deviceID to {\ttfamily d2} variable (line 16) and then stores it into the {\ttfamily outState} bundle (line 21). (4) Upon rotation of the device, the activity is destroyed and recreated. Upon destroying, {\ttfamily onSaveInstanceState()} is invoked which stores the deviceID in the {\ttfamily outState} bundle again. Three callbacks ({\ttfamily onCreate()}, {\ttfamily onRestoreInstanceState()}, and {\ttfamily onResume()}) are invoked upon recreating the activity. This callback sequence obtains deviceID from the {\ttfamily state} object (line 9) and leaks it by sending a text message (line 13) to a number given in {\ttfamily recpNo} variable.

\definecolor{mygreen}{rgb}{0,0.6,0}
\definecolor{mygray}{rgb}{0.5,0.5,0.5}
\definecolor{mymauve}{rgb}{0.58,0,0.82}

\lstset{language=Java, basicstyle=\scriptsize\ttfamily , numbers=left, stepnumber=1, numbersep=5pt,  keywordstyle=\color{blue},   commentstyle=\color{mygreen}, 
frame=single, showspaces=false, numberstyle=\tiny\color{black}, rulecolor=\color{black}, stringstyle=\color{mymauve},
showstringspaces=false, showtabs=false, tabsize=2,
 breaklines=true, breakatwhitespace=false, caption={Motivating Example} , label={listing1}}

\noindent\begin{minipage}{14cm}  
\begin{lstlisting}
public class MainActivity extends Activity {
	String d1 = ""; String d2 = ""; 
	String d3 = ""; String recpNo = "1066156686";

	void onCreate(Bundle instance){
		setContentView(R.layout.activity_main);
	}
	void onRestoreInstanceState(Bundle state){
		this.d3 = state.getString("myData");
	}
	void onResume(){
		if(!d3.equals(""))
			SmsManager.getDefault().sendTextMessage(recpNo, null, d3, null, null);
	}
	void onUserLeaveHint(){
		this.d2 = this.d1;
		TelephonyManager tMgr = (TelephonyManager) getApplicationContext().getSystemService(TELEPHONY_SERVICE);
		this.d1 = tMgr.getDeviceId();
	}
	void onSaveInstanceState(Bundle outState){
		outState.putString("myData", this.d2);
	}
	void onBtnClicked(View v){
		if(v.getId() == R.id.add){ ... }
		...      //calculator code here.
	}	
}
\end{lstlisting}
\end{minipage} 

The above attack leaks sensitive information when a set of callbacks are invoked in a specific order. The {\ttfamily onUserLeaveHint()} callback is invoked twice by two different events to obtain and store the deviceID. While these two events can occur in any order, they must be followed by a device-rotation event to successfully launch the attack. These events can be triggered easily in real-world scenarios because pressing Back or Home button is commonly used to switch between the activities and the apps. However, their invoked callback sequences must be analyzed in the given order to detect this attack. Moreover, the example involves callbacks such as {\ttfamily onUserLeaveHint()} and {\ttfamily onSaveInstanceState()} to launch the attack. The system invokes these callbacks in response to different events upon activity. Android-supplied activity life cycle model omits these callbacks and their associated states, and hence, it will not be able to detect the above attack. Therefore, a better activity life cycle model is required.

\section{Dexteroid Framework}

\begin{figure}[ht!]
\centering
\includegraphics[width=120mm]{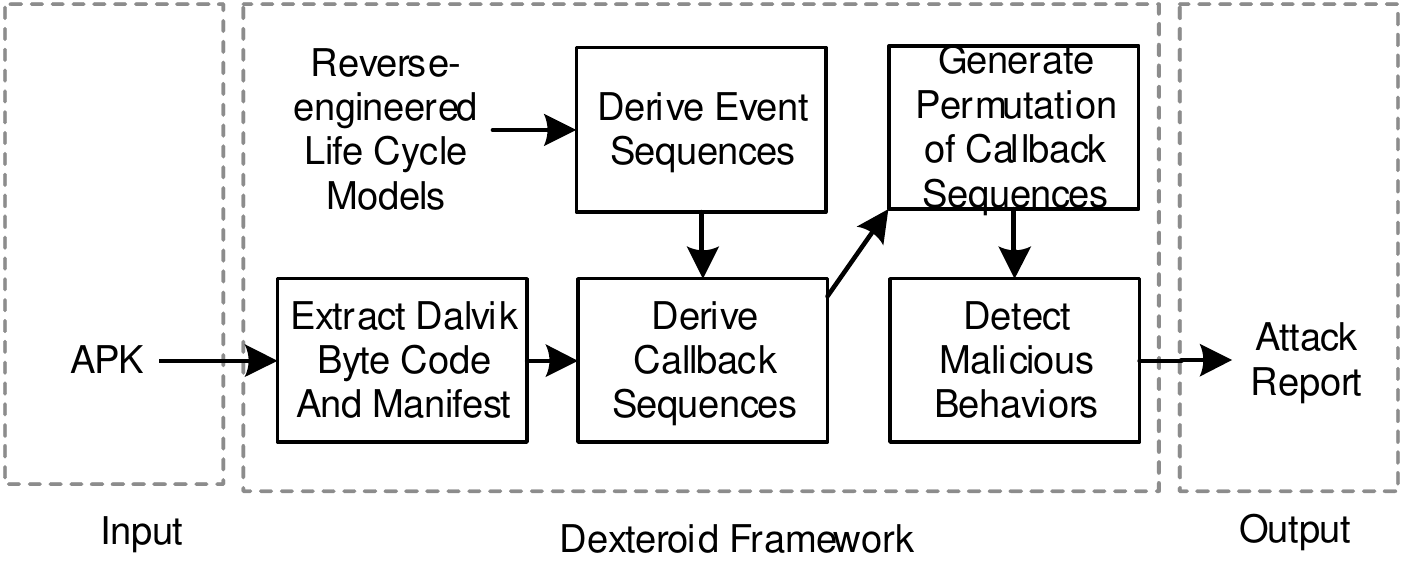}
\caption{Dexteroid Framework}
\label{framework}
\end{figure}

This section presents the Dexteroid framework, as shown in Figure \ref{framework}. It uses reverse-engineered life cycle models that include states and transitions previously omitted in Android-supplied life cycle models. Dexteroid inputs an Android Application Package (APK) file and decompiles it using Androguard \cite{Androguard} to extract Dalvik bytecode and its manifest file. It identifies registered components from the manifest file and analyzes these components iteratively. For each component, the framework systematically derives event sequences from the reverse-engineered life cycle model and then derives callback sequences from these event sequences. These callback sequences are guaranteed to be feasible because they are derived from the event sequences which themselves are derived from the component life cycle model. Dexteroid then generates permutations of callback sequences and performs taint analysis on them to detect malicious behaviors. The following sections describe the reverse-engineering of these life cycle models, and derivation of callback sequences in detail.

\subsection {Reverse-Engineering of Activity Life Cycle Model}  \label{activityLCM}

We follow \cite{Franke2011} to reverse-engineer the activity life cycle model. An Android app is developed in which each activity contains all life cycle callbacks, extracted from Android API documentation. We add {\ttfamily Log.v()} API call in all the callbacks of the activity. When a callback is invoked, the API call records that information. An extensive set of operations are performed upon an activity to trigger events which an activity can possibly receive during its execution. Table \ref{activityTable} shows a subset of user and system operations which we use to trigger events upon an activity. When an event is triggered, a set of life cycle callbacks are invoked in a specific order. For example, when a user opens an activity for the first time, createActivity event is triggered. This event causes the system to invoke a sequence of callbacks in the activity as shown in Table \ref{activityTable}. Using triggered events and the observed activity behavior, we reconstruct activity life cycle model as shown in Figure \ref{activitySM}. We use UML state diagram with composite sequential states (CSS) \cite{BoochUML} to present the proposed model and refer to it as the CSS model (of an Android activity). This model contains new states (e.g., PostCreated) as well as all the states (e.g., Created) shown in the Figure \ref{coreLifeCycle}.

\begin{ThreePartTable}

\begin{scriptsize}
\begin{longtable}{P{3.2cm} P{1.5cm} P{5.9cm}}  
\caption[User or System Operations and Resultant Triggered Events and Callbacks]{User or System Operations and Resultant Triggered Events and Callbacks} \label{activityTable} \\

\hline \hline {\multirow{2}{*}{User or System Operation}} & Triggered Event$^{a}$ & {\multirow{2}{*}{Invoked Activity Life Cycle Callbacks}} \\ \hline \hline 
\endfirsthead

\multicolumn{3}{c}%
{{\bfseries \tablename\ \thetable{} -- continued from previous page}} \\
\hline \hline {\multirow{2}{*}{User or System Operation}} &
Triggered Event$^{a}$ &
{\multirow{2}{*}{Invoked Activity Life Cycle Callbacks}} \\ \hline \hline 
\endhead

\multicolumn{3}{|r|}{{Continued on next page}} \\ \hline
\endfoot
\hline \hline 
\endlastfoot
1) Tap on app icon or start from another activity & createActivity${^\ast}$  & {\ttfamily onCreate()}, {\ttfamily onStart()}, {\ttfamily onPostCreate()}, {\ttfamily onResume()}, {\ttfamily onPostResume()}\\ \hline
(2) After (1), press Back button on the device & backPress & {\ttfamily onPause()}, {\ttfamily onStop()}, {\ttfamily onDestroy()}\\ \hline
(3) After (2), tap on app icon to start activity & createActivity${^\ast}$ & {\ttfamily onCreate()}, {\ttfamily onStart()}, {\ttfamily onPostCreate()}, {\ttfamily onResume()}, {\ttfamily onPostResume()}\\ \hline
(4) After (1), rotate the device (e.g., from vertical to horizontal) & confPR & {\ttfamily onPause()}, {\ttfamily onSaveInstanceState()},  {\ttfamily onStop()}, {\ttfamily onDestroy()}, {\ttfamily onCreate()}, {\ttfamily onStart()}, {\ttfamily onRestoreInstanceState()}, {\ttfamily onPostCreate()}, {\ttfamily onResume()}, {\ttfamily onPostResume()}\\ \hline
(5) After (1), alarm goes off & stopActivity${^\star}$ & {\ttfamily onPause()}, {\ttfamily onCreateDescription()}, {\ttfamily onSaveInstanceState()}, {\ttfamily onStop()}\\ \hline
(6) After (5), snooze or stop the alarm & restartActivity${^\dagger}$ & {\ttfamily onRestart()}, {\ttfamily onStart()}, {\ttfamily onResume()}, {\ttfamily onPostResume()}\\ \hline
(7) After (5), rotate device and snooze or stop the alarm & confPOS${^\ddagger}$ & {\ttfamily onDestroy()}, {\ttfamily onCreate()}, {\ttfamily onStart()}, {\ttfamily onRestoreInstanceState()}, {\ttfamily onPostCreate()}, {\ttfamily onResume()}, {\ttfamily onPostResume()}\\ \hline
(8) After (1), press Home button on phone & overlapActivity${^\circ}$ & {\ttfamily onUserLeaveHint()}, {\ttfamily onPause()}, {\ttfamily onCreateDescription()}, {\ttfamily onSaveInstanceState()}, {\ttfamily onStop()}\\ \hline
(9) After (8), start the app from launcher menu & restartActivity${^\dagger}$ & {\ttfamily onRestart()}, {\ttfamily onStart()}, {\ttfamily onResume()}, {\ttfamily onPostResume()}\\ \hline
(10) After (8), rotate device and start the app from launcher menu & confPOS${^\ddagger}$ & {\ttfamily onDestroy()}, {\ttfamily onCreate()}, {\ttfamily onStart()}, {\ttfamily onRestoreInstanceState()}, {\ttfamily onPostCreate()}, {\ttfamily onResume()}, {\ttfamily onPostResume()}\\ \hline
(11) After (8) or (5), the system kills the process to recover resources & killProcess${^\bullet}$ & {\ttfamily onDestroy()}\\ \hline
(12) After (1), tap on add to bookmark icon on the device (e.g., Nexus 5 phone) & overlapActivity${^\circ}$ & {\ttfamily onUserLeaveHint()}, {\ttfamily onPause()}, {\ttfamily onCreateDescription()}, {\ttfamily onSaveInstanceState()}, {\ttfamily onStop()}\\ \hline
(13) After (12), remove the app from bookmarks list & killProcess${^\bullet}$ & {\ttfamily onDestroy()}\\ \hline
(14) After (1), start a dialog activity from this activity & hideActivityPartially${^\diamond}$ & {\ttfamily onUserLeaveHint()}, {\ttfamily onPause()}, {\ttfamily onCreateDescription()}, {\ttfamily onSaveInstanceState()}\\ \hline
(15) After (14), press Back button & gotoActivity${^\triangle}$ & {\ttfamily onResume()}, {\ttfamily onPostResume()} \\ \hline
(16) After (14), press Home button & savStop${^\wedge}$ & {\ttfamily onStop()}\\ \hline
(17) After (16), start the app from launcher menu & savRestart & {\ttfamily onRestart()}, {\ttfamily onStart()}\\ \hline
(18) After (17), press Home button & savStop${^\wedge}$ & {\ttfamily onStop()}\\ \hline
(19) After (17), press Back button & gotoActivity${^\triangle}$ & {\ttfamily onResume()}, {\ttfamily onPostResume()}\\ \hline
(20) After (14) or (17) , rotate the device & confSTP & {\ttfamily onStop()}, {\ttfamily onDestroy()}, {\ttfamily onCreate()}, {\ttfamily onStart()}, {\ttfamily onRestoreInstanceState()}, {\ttfamily onPostCreate()}, {\ttfamily onResume()}, {\ttfamily onPostResume()}, {\ttfamily onPause()}\\ \hline 
(21) After (20), press Back button & gotoActivity${^\triangle}$ & {\ttfamily onResume()}, {\ttfamily onPostResume()}\\ \hline
(22) After (20), press Home button & gotoStop & {\ttfamily onCreateDescription()}, {\ttfamily onSaveInstanceState()}, {\ttfamily onStop()}\\ \hline
(23) After (20), rotate the device & confPAU & {\ttfamily onSaveInstanceState()}, {\ttfamily onStop()}, {\ttfamily onDestroy()}, {\ttfamily onCreate()}, {\ttfamily onStart()}, {\ttfamily onRestoreInstanceState()}, {\ttfamily onPostCreate()}, {\ttfamily onResume()}, {\ttfamily onPostResume()}, {\ttfamily onPause()}\\ \hline
(24) After (22), rotate device and start the app & confSTO & {\ttfamily onDestroy()}, {\ttfamily onCreate()}, {\ttfamily onStart()}, {\ttfamily onRestoreInstanceState()}, {\ttfamily onPostCreate()},  {\ttfamily onResume()}, {\ttfamily onPostResume()}, {\ttfamily onPause()}\\ \hline
(25) After (13), open the app & createActivity${^\ast}$ & {\ttfamily onCreate()}, {\ttfamily onStart()}, {\ttfamily onPostCreate()}, {\ttfamily onResume()}, {\ttfamily onPostResume()}\\ \hline
(26) After (1), start another activity & overlapActivity${^\circ}$ & {\ttfamily onUserLeaveHint()}, {\ttfamily onPause()}, {\ttfamily onCreateDescription()}, {\ttfamily onSaveInstanceState()}, {\ttfamily onStop()}\\ \hline
(27) After (1), open notification bar and tap on a notification & overlapActivity${^\circ}$ & {\ttfamily onUserLeaveHint()}, {\ttfamily onPause()}, {\ttfamily onCreateDescription()}, {\ttfamily onSaveInstanceState()}, {\ttfamily onStop()} \\ \hline
(28) After (1), lock phone by pressing Power button & stopActivity${^\star}$ & {\ttfamily onPause()}, {\ttfamily onCreateDescription()}, {\ttfamily onSaveInstanceState()}, {\ttfamily onStop()}\\ \hline
(29) After (28), unlock the device & restartActivity${^\dagger}$ & {\ttfamily onRestart()}, {\ttfamily onStart()}, {\ttfamily onResume()}, {\ttfamily onPostResume()}\\ \hline
(30) After (1), receive the phone call & stopActivity${^\star}$ & {\ttfamily onPause()}, {\ttfamily onCreateDescription()}, {\ttfamily onSaveInstanceState()}, {\ttfamily onStop()}\\ \hline
(31) After (30), accept and finish the call or reject the call & restartActivity${^\dagger}$ & {\ttfamily onRestart()}, {\ttfamily onStart()}, {\ttfamily onResume()}, {\ttfamily onPostResume()}\\ \hline
(32) After (1), message preview notification is received (e.g., from Viber \cite{viber} app) & hideActivityPartially${^\diamond}$ & {\ttfamily onUserLeaveHint()}, {\ttfamily onPause()}, {\ttfamily onCreateDescription()}, {\ttfamily onSaveInstanceState()} 
\end{longtable}
  \begin{tablenotes}
    \item[$^{a}$]{Different user or system operations can trigger the same event---marked with the same symbol---in the above table. For example, user operations (8), (12), (26), and (27) trigger overlapActivity event and invoke the same callback sequence for the activity.}
    \end{tablenotes}
\end{scriptsize}    
\end{ThreePartTable}

\newcommand*\mycircle{\textcircled{\textbullet}}

\newcommand\bulletarrow{{\ooalign{\raise0.01ex\hbox{\scalebox{2}{$\bullet$}}}}}

\subsection{Our Proposed CSS Model}  \label{cssModel}

This section describes elements of the CSS model and its interpretation in detail. 

\begin{figure*}
\centering
\includegraphics[width=140mm]{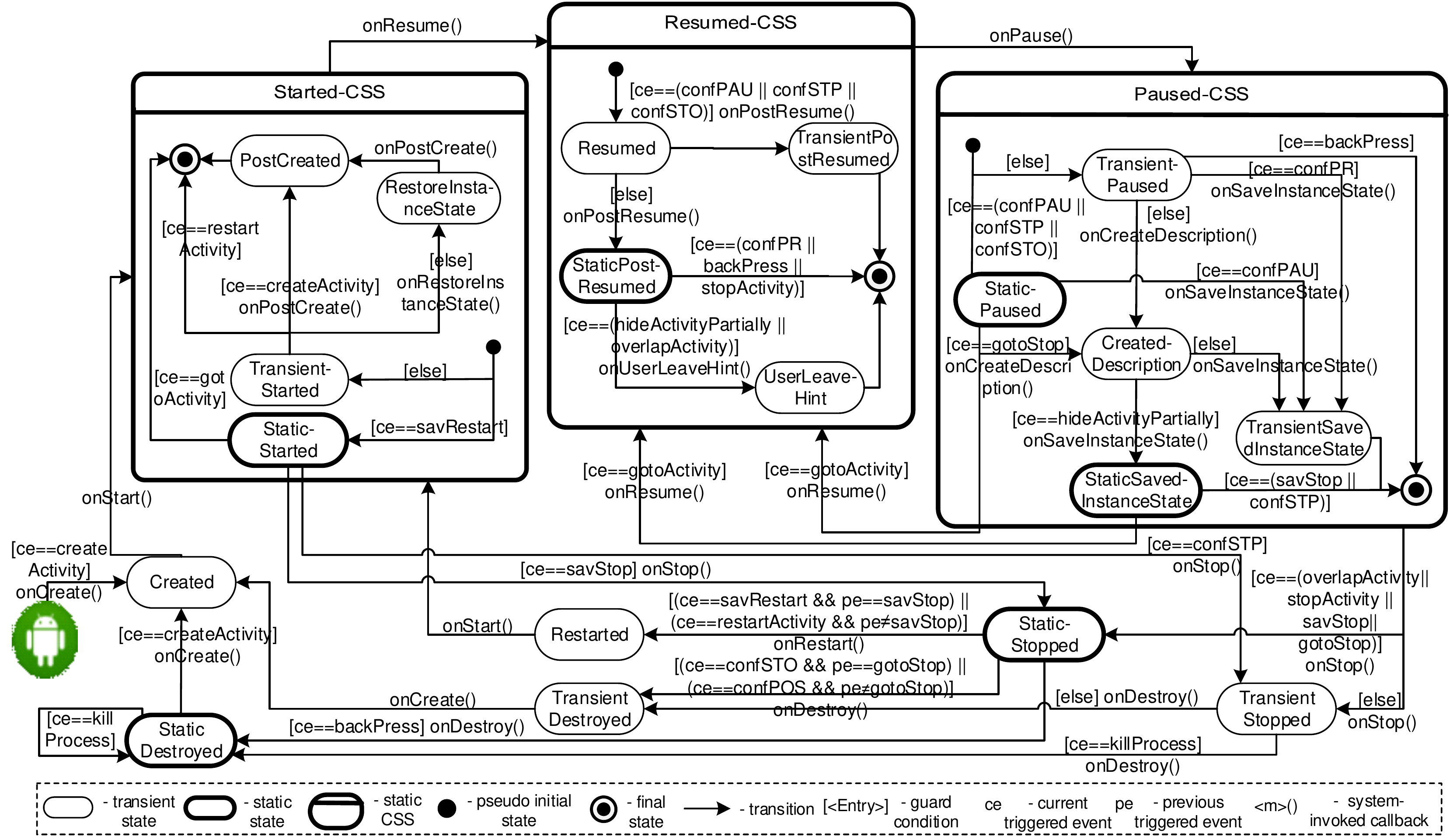}
\caption{Composite Sequential State Model (The Reverse-engineered Activity Life Cycle Model}
\label{activitySM}
\end{figure*}

{\noindent \bf Composite Sequential State:} A CSS is a region of states which are related by state transitions and only one of its states can be active at any given time \cite{BoochUML}. When a CSS is entered, the activity makes a transition from its pseudo-initial state (represented with \bulletarrow) to the default state automatically. Similarly, a final state (represented with \mycircle) shows that no more state transitions are possible within this CSS. A CSS is a static CSS if any of the substates is static, otherwise it is a transient CSS. In Figure \ref{activitySM}, three life cycle states (Started, Resumed and Paused) are decomposed into their respective CSS states (Started-CSS, Resumed-CSS and Paused-CSS). At a higher level, the activity shows similar behavior both in a CSS and its counterpart non-CSS state but CSS contains additional states to characterize more detailed behavior of the activity.

{\noindent \bf Guard Conditions:} An activity can receive different types of events in a static state. Based on the current and/or previously triggered events, the activity can make a specific sequence of state transitions. The event-related information required to make a transition is shown using guard conditions, enclosed in `[' and `]' in Figure \ref{activitySM}. A transition from a static state requires new event to be triggered. Guard condition for such a transition indicates the new (and now current) event which needs to be triggered for this transition. However, transition from a transient state does not require new event to be triggered from this state. Guard condition for such a transition is evaluated for the current triggered event from the last-visited static state. The activity makes a transition when guard condition is true or no guard condition is shown (which means always true). In a case, when a transition can occur in response to multiple events, it is represented by `[else]' branch, which is evaluated after all \emph{if} conditions have been evaluated to false. 

{\noindent \bf Dual State Behavior:} An activity stays only in a static state. However, depending upon the triggered event, the static state may also behave like a transient state. In that case, the activity visits the static state and moves to the next state immediately. To distinguish between specific behaviors of a state, we present such static state as two states. For instance, Stopped state which shows dual state behavior is shown as TransientStopped state and StaticStopped state in Figure \ref{activitySM}. Both states have the same callback method ({\ttfamily onStop()}) for the activity to reach these states. 

{\noindent \bf Explanation:} All of the CSS states shown in Figure \ref{activitySM} are static CSS. The Resumed-CSS state is reached with a call to {\ttfamily onResume()} callback method and it contains Resumed, StaticPostResumed, TransientPostResumed and UserLeaveHint states. When createActivity event is triggered from the initial state (Android robot icon), the activity goes through different states to reach Resumed-CSS state. Upon entrance to this state, the activity makes a transition from its pseudo-initial state (\bulletarrow) to Resumed state automatically. Depending upon the triggered event, it moves to StaticPostResumed state or TransientPostResumed state. When reached its StaticPostResumed state, the activity waits for user-interaction in this state and requires a trigger to make next transition. This indicates that Resumed state itself is a transient state as opposed to what is shown in Figure \ref{coreLifeCycle} while StaticPostResumed state is a static state. 

A triggered event in a static state can make activity go through a sequence of transient states until it reaches its next static state. The specific sequence of states that it goes through is dependent upon the triggered event. For example, rotating a device from portrait view to landscape view in StaticPostResumed state triggers a confPR event ((4) in Table \ref{activityTable}). This event indicates the change in configuration for the current app and the system destroys current activity in response and recreates it in accord with the new configuration. The activity visits a sequence of transient states in the CSS model to reach back to its StaticPostResumed state. Specifically, it visits TransientPaused, TransientSavedInstanceState, TransientStopped, TransientDestroyed, Created, TransientStarted, RestoredInstanceState, PostCreated, Resumed, and StaticPostResumed states. In StaticPostResumed state, the user can trigger the same or other event regardless of the previously triggered event. However, in some case, the next state transition from a static state may be dependent upon the previously triggered event from a static state. This scenario is shown with additional guard conditions (\&\&) in Figure \ref{activitySM}. For example, consider the state transition from StaticStopped state to Restarted state under the same operation: the savRestart event is triggered only if the previously triggered event is savStop, otherwise the restartActivity event is triggered. Though both events bring activity to Restarted state but the future transitions between the transient activity states are dependent upon the triggered event. The activity visits Restarted and StaticStarted states for savRestart while it visits Restarted, TransientStarted, Resumed, and StaticPostResumed states for the restartActivity event.

Table \ref{activityTable} shows that one single event can be triggered by many different operations. For example, user operations (8), (12), (26), and (27) trigger overlapActivity event and cause the system to invoke the same callback sequence for the activity. Thus, Figure \ref{activitySM} shows only one triggered event from StaticPostResumed state for all these user operations. We perform reverse-engineering on Android 4.4 (alias KitKat) version which is the most widely used Android platform installed by the users\footnote{Around 40\% of the users are using Android KitKat at the time of publication (\url{https://developer.android.com/about/dashboards/index.html}).}. The reverse-engineered life cycle model in Figure \ref{activitySM} is much closer (if not completely same) to the real life cycle than the Android-supplied activity life cycle model. This is due to use of an extensive set of input events, which cover most of the events that can be triggered in day-to-day app usage. This results in high coverage of callback flows in the activity life cycle. However, like all other dynamic analysis approaches, our method may miss some callback flows present in the real activity life cycle. In the future, we will define test-adequacy code coverage criteria, and utilize automatic testing tools as well as code coverage tools to substantially increase code coverage of the activity life cycle.

{\noindent \bf Example:} The motivating example contains four operations. The first operation triggers createActivity event and the activity reaches its StaticPostResumed state. The second operation triggers hideActivityPartially and gotoActivity events and in response, the activity moves to StaticSaveInstanceState and then StaticPostResumed state. The third operation triggers overlapActivity and gotoActivity events while the fourth operation triggers confPR event. The activity visits many states against these events before it reaches StaticPostResumed state.

\subsection {Deriving Event Sequences} \label{secMalBhvrDetction}

A motivated attacker can design a targeted attack which is launched only by a specific ordering of the life cycle callback methods. Existing techniques such as \cite{DroidSafe} analyze all possible orderings of callbacks to detect such attacks. However, such an approach may lead to many false positives because most of the callback sequences will not occur in a real Android app. For example, given an activity that contains all callbacks given in the CSS model, an ordering sequence containing {\ttfamily onCreate()} right after {\ttfamily onStart()} would violate the activity life cycle model and cannot occur in an app.

\subsubsection {Event Sequence-based Analysis} An activity waits for user interaction in its StaticPostResumed state. It can receive different events in this sate. A triggered event can cause the activity to visit a specific sequence of states. Following a visited state sequence in the CSS model, one can derive callback sequence that the system invokes for a particular event. Thus, performing analysis at event level becomes promising to analyze activity component because all event-generated callback sequences follow activity life cycle model and can occur in an Android app. To detect attacks that are launched by specific ordering of events, an analysis tool can consider permutation of events (i.e., their generated callback sequences) for analysis. However, such an approach can lead to false positives because some of the events in the CSS model occur only in a specific order. For example, in a StaticPostResumed state, a user tap on Home button of the device triggers overlapActivity event ($e_1$) which moves current activity to its StaticStopped state and hides the activity and the app into background. The user cannot interact with the activity in this state unless an event is triggered which brings activity back to its StaticPostResumed state. Thus, reopening the app triggers restartActivity event ($e_2$) and the activity reaches its StaticPostResumed state. These two events ($e_1$ and $e_2$) can occur only in this specific order. Similarly, other events such as hideActivityPartially and gotoActivity can be triggered only in the given order. Thus, analysis of event permutations can produce false positives because some event sequences generated by the event permutation may not occur in an app.

A key observation from the above discussion is that the event sequence ($e_1$-$e_2$) makes the activity visit a combined sequence of states which starts from, and ends with StaticPostResumed state. The user cannot directly interact with the activity (e.g., on UI elements) in the intermediate states except triggering an $e_2$ event. Such an event sequence behaves like an individual event. Before and after execution of the event sequence, the activity stays in its StaticPostResumed state where it can receive new events, independent of previously triggered events. One can derive such event sequences from the CSS model which implies that the order of events in the sequences is guaranteed to be feasible. Thus, considering event sequences in place of individual events becomes very promising to accurately analyze activity component in Android apps. The following section presents an algorithm to derive such event sequences from the CSS model.  

\algnewcommand\algorithmicinput{\textbf{Input:}}
\algnewcommand\INPUT{\item[\algorithmicinput]}

\def\therule{\makebox[\algorithmicindent][l]{\hspace*{.5em}\vrule height .75\baselineskip depth .25\baselineskip}}%

\makeatother
\setlength{\algorithmwidth}{.9\textwidth}

\begin{algorithm}[ht]
 \begin{algorithmic}[1]
\INPUT 
\Statex $M$ \Comment State transition model
\Procedure{ActivityEventSequences} {$M$}
	\State $s$ = $M$.getInitialState()  \Comment{$s$ gets initial state}
	\State $e$ = NIL \Comment{$e$ gets current triggered event}
	\State $z$ = StaticPostResumed \Comment{$z$ gets a goal state}
	\State $T$ = NIL \Comment{$T$ gets the running event sequence}
	\ForAll {State $q$ $\in$ $M$}  
		\State $q$.$color$ = WHITE
	\EndFor
	\State {\Call{EventSequenceDerivation}{$M, s, e, z,T$}}
\EndProcedure
  \end{algorithmic}
 \caption{Driving Event Sequences from the Activity Life Cycle Model}\label{eventDerivationActivity}
\end{algorithm}

\subsubsection{Event Sequence Derivation Algorithm}

The CSS model shown in Figure \ref{activitySM} is a compact representation of the activity life cycle model. To facilitate the formulation and implementation of the event sequence derivation algorithm, the CSS model is converted into a flat state diagram using the semantics presented in Section \ref{cssModel}. The resulting state transition model $M$ retains the static and transient states, while eliminates the CSS states. In addition, a transition with a disjunctive guard condition "[$e_1$ || $e_2$ || ... || $e_n$]" is replaced by $n$ transitions, which can be triggered by the $n$ events, respectively. 

More formally, the model $M$ is a 5-tuple ($S$, $S_0$, $\Sigma$, $\Lambda$, $\delta$):

\begin{itemize}
\item $S$ is a set of static and transient states (e.g., Created state)
\item $S_0$ is an initial state (Android robot)
\item $\Sigma$ is a set of input events (e.g., createActivity)
\item $\Lambda$ is a set of callbacks (e.g., {\ttfamily onCreate()})
\item $\delta$: $S$ $\times$ $\Sigma$ $\rightarrow$ $S$ $\times$ $\Lambda$ is a state transition function, which returns the next state and a callback function. 
\end{itemize}

Algorithm \ref{eventDerivation} describes an algorithm for deriving event sequences from state model $M$ by extending Depth First Search (DFS) algorithm described in \cite{Cormen}. It explores all the events which can be triggered from a given static state and makes state transitions based upon triggered events. It builds event sequences along the way which make activity reach its goal state (i.e., StaticPostResumed for the CSS model). The algorithm is passed five parameters: (1) a state model $M$, (2) current state $s$, (3) current triggered event $e$ (guard condition based on which next state transition is determined), (4) goal state $z$, and (5) the running event sequence $T$. 

Algorithm \ref{eventDerivationActivity} takes the state model $M$ as an input and performs initializations. It then invokes Algorithm \ref{eventDerivation} to generate event sequences for activity life cycle model. Starting from the initial state, Android robot (which is a static state), the algorithm explores all possible transitions against different events which can occur from a static state (line 14). In each iteration of the loop, a transition against a specific event is extracted and that event information is passed to future invocations of the algorithm. The algorithm traverses a sequence of transient states based on the triggered event (line 22) until it reaches the next static state. The algorithm returns when it reaches the StaticPostResumed state because this state is already being explored in earlier invocation of the algorithm.

\setlength{\algorithmwidth}{.95\textwidth}

\begin{algorithm}[ht!]
 \begin{algorithmic}[1]
 \INPUT 
\Statex $M$ \Comment State transition model
\Statex $s$ \Comment  Current state
\Statex $e$ \Comment Current triggered event
\Statex $z$ \Comment Goal state
\Statex $T$ \Comment Running event sequence
\Procedure{EventSequenceDerivation} {$M, s, e, z,T$}
	\If{$s$.$color$ $\neq$ WHITE $\&\&$ $s$.$name$ $==$ $z$}
		\State print $T$  \Comment{$T$ is an event sequence, starting from createActivity}
		\State \Return
	\ElsIf {$s$.$color$ $==$ RED $\&\&$ $s$.$type$ $==$ STATIC}
		\State \Return
	\Else
		\If{$s$.$type$ $==$ STATIC}
			\State currStateColor = $s$.$color$
			\If {$s$.$color$ $==$ WHITE}
			    \State{$s$.$color$ $=$ GREY}
			\Else 
			    \State{$s$.$color$ $=$ RED}
			\EndIf    
			\ForAll {Transition $tr$ $\in$ $M$.nextTransitions($s$, $e$)}  
				\State $n$ $=$ $M$.getTriggeredEvent($tr$,$e$)
				\State $s'$ $=$ $M$.getDestination($tr$)
				\State $tempT$ $=$ $T$
				\If{$s' \neq s$} 	\Comment{avoid self-loop}
					\State Append $n$ to $tempT$ 
					\State{\Call{EventSequenceDerivation}{$M,s',n,z,tempT$}}
				\EndIf	
			\EndFor
		       \State $s$.$color$ $=$ currStateColor
		\ElsIf {$s$.$type$ $==$ TRANSIENT}
			\State $tr$ $=$ $M$.getTransition($s$,$e$) 
			\State $s'$ $=$ $M$.getDestination($tr$)
			\If{$s' \neq s$}  \Comment{avoid self-loop}
				\State{\Call{EventSequenceDerivation}{$M,s',e,z,T$}}
			\EndIf
		\EndIf
	\EndIf 	
\EndProcedure

  \end{algorithmic}
 \caption{Event Sequence Derivation From A State Model}
 \label{eventDerivation}
\end{algorithm}

{\noindent \bf{Handling Loops:}} The $color$ attribute for a state is introduced to show its visited status by the algorithm. Initially, all the states are set to WHITE. When events are explored from a static state for first time, its $color$ is set to GREY; upon another visit, its $color$ is set to RED. The three values for $color$ attribute are introduced to capture all possible flows (at least once) among the activity states. For example, the state transitions between StaticStopped and StaticStarted states form a loop in Figure \ref{activitySM} and it is desired to capture the flow from end of loop to start of the loop at least once. Upon first visit, the $color$ of StaticStopped state (start of the loop) is set to GREY and savRestart event is explored. The activity reaches its StaticStarted state. A savRestart event from this state brings activity back to StaticStopped state. The $color$ of StaticStopped is set to RED and all events are explored again but the algorithm returns if StaticStopped is seen again (line 5-6). The algorithm ignores self-loops while making state transitions between the states. The $color$ of transient states is not set or used because they do not introduce new events to the sequences and just make transitions based on an already triggered event. 

\begin{figure}[ht]
\centering
\includegraphics[width=0.90\textwidth]{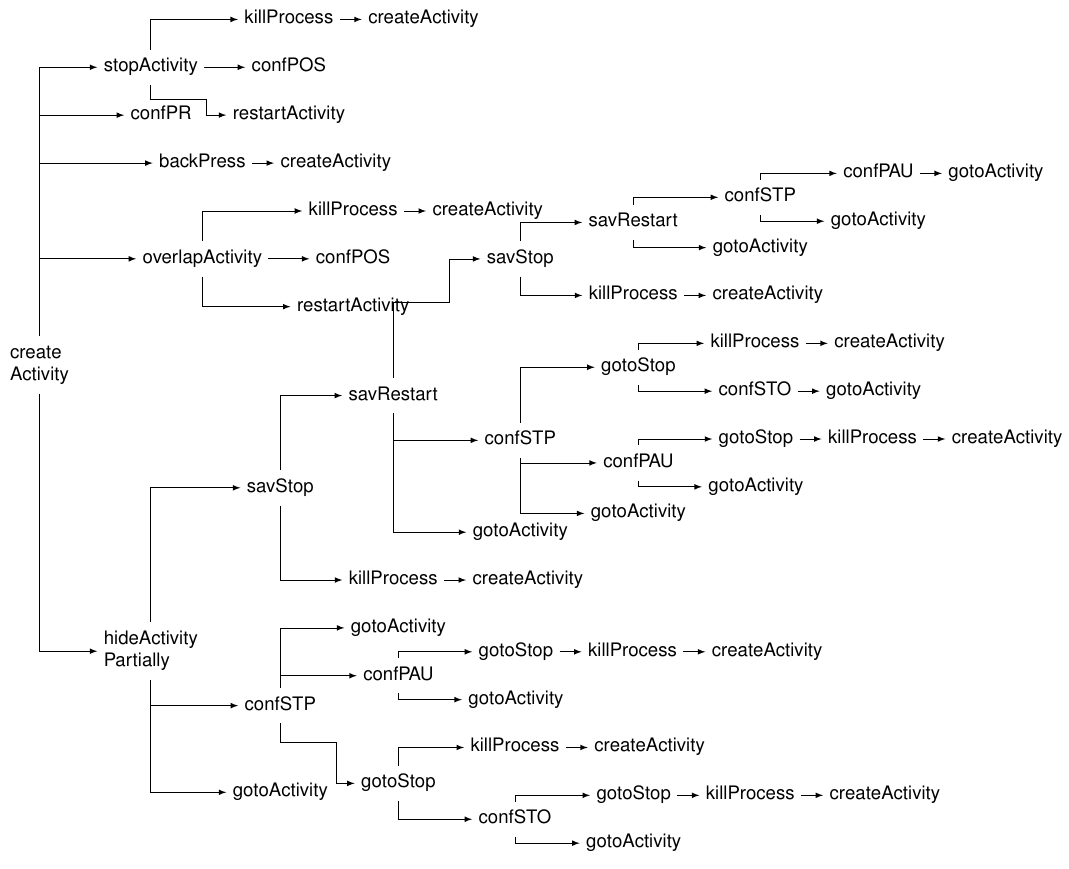}
\caption{Twenty-six derived event sequences for the reverse-engineered activity life cycle model}
\label{activityEvents}
\end{figure}

{\noindent \bf{Derived Event Sequences:}}  All event sequences produced by the algorithm are shown in Figure \ref{activityEvents}. When createActivity event is triggered for the first time, the activity reaches its StaticPostResumed state and can then follow any of the 26 event sequences presented in Figure \ref{activityEvents}. After the createActivity event, all other events in a sequence move the activity from StaticPostResumed state and bring it back to StaticPostResumed state.

\tikzset{>=latex}

\forestset{
  stepwise/.style n args=2{
    edge path={
      \noexpand\path [draw, \forestoption{edge}] (!u.parent anchor) |- +(#1,#2) |- (.child anchor)\forestoption{edge label};
    }
  },
  my shading/.style={
    for tree={
      text/.wrap pgfmath arg={black!##1!#1}{10*level()},
      edge/.wrap pgfmath arg={->, draw=black!##1!#1}{10*level()},
    },
  },
}

\subsection{Service Life Cycle Model} \label{serviceLCM}

\subsubsection{Android-Supplied Service Life Cycle Model}

A service performs long running operations in the background and does not provide a user interface. The life cycle of a service \cite{service} depends upon how a service is run. A service started by another application component (e.g., activity) using {\ttfamily startService()} can run indefinitely, even if the component itself is destroyed later. The service is in Started state now and should stop itself after the desired operations are completed. On the other hand, a service bound to another component using {\ttfamily bindService()} runs as long as the component is bound to it. The service can be bound to many components simultaneously but it is unbound when any of the binding components calls {\ttfamily unbindService()} API. 

\subsubsection{The Reverse-Engineered Service Life Cycle Model}

\begin{figure*}[ht!]
\centering
\includegraphics[width=120mm]{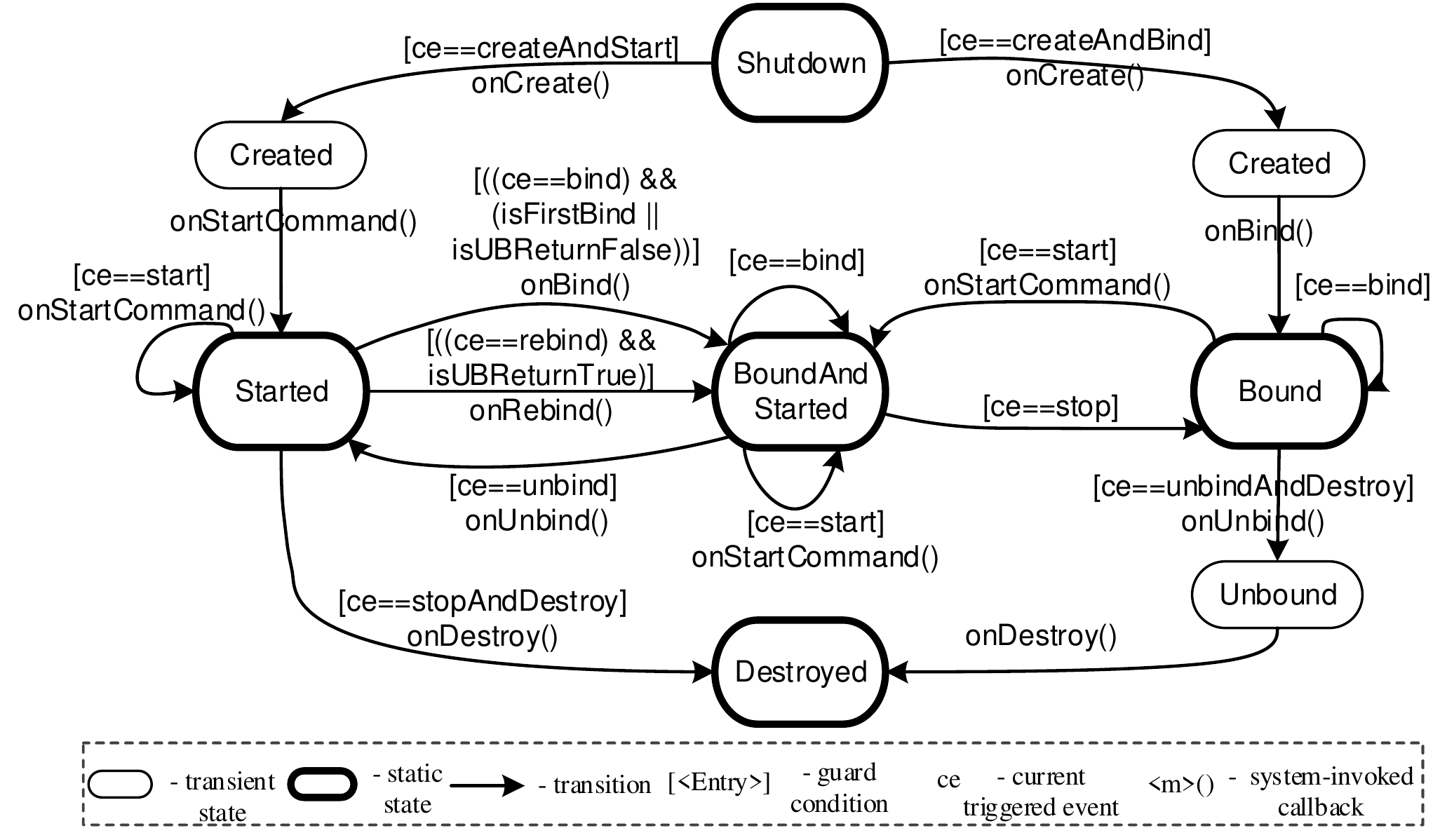}
\caption{The Reverse-Engineered Service Life Cycle Model}
\label{serviceSM}
\end{figure*}

Like activity, the Android-supplied service life cycle model omits some states and transitions between the states. We apply the above mentioned technique (also described in \cite{serviceLCMFrank}) to reverse-engineer the service life cycle model. Static and transient states are used to model the service life cycle as shown in Figure \ref{serviceSM}. The service is initially in Shutdown state (static state). With events triggered from this state, the service can reach its Started or Bound state. However, the service can be bound as well as started simultaneously to make it reach its BoundAndStarted state. In a Started state, if a service is being bound for the first time or after the {\ttfamily onUnbind()} invocation has returned false, the system invokes {\ttfamily onBind()} callback otherwise it invokes {\ttfamily onRebind()} callback. This has been shown using guard conditions in Figure \ref{serviceSM}. The other API calls such as {\ttfamily startForeground()} are not shown in Figure \ref{serviceSM} because they do not cause any interesting state transition or callback invocation for the security analysis purpose. The completeness verification of the life cycle model is left as part of our future work.

\subsubsection{Driving Event Sequences}

\begin{figure}[ht]
\centering
\includegraphics[width=0.90\textwidth]{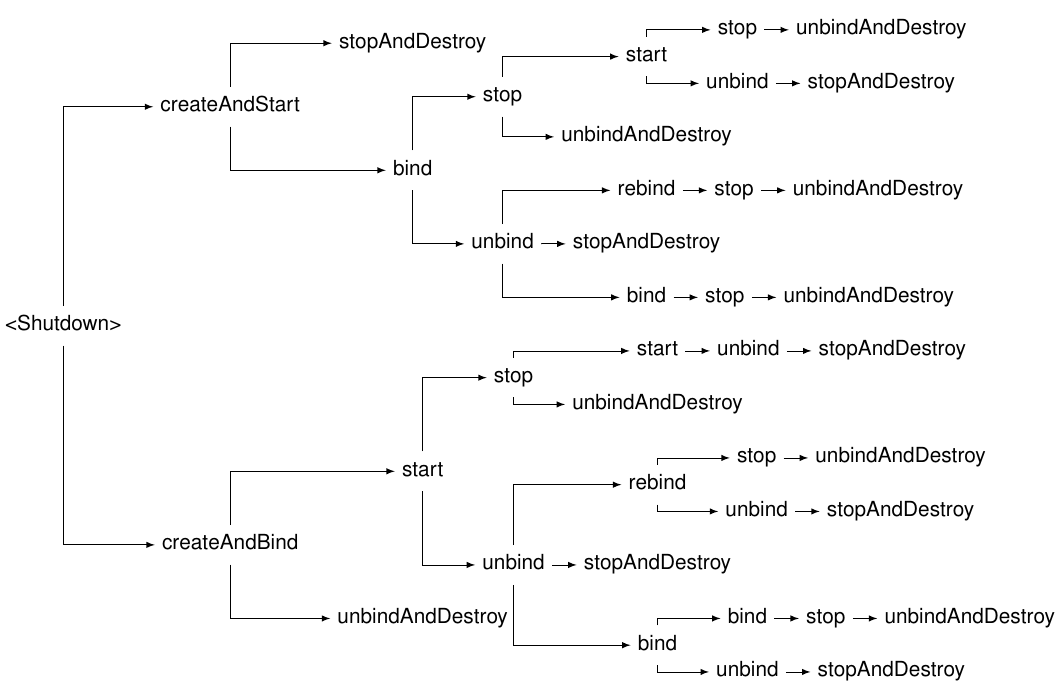}
\caption{15 Derived Event Sequences for the Reverse-Engineered Service Life Cycle Model}
\label{serviceEvents}
\end{figure}

Algorithm \ref{eventDerivation} derives event sequences from a state transition model which is based on static and transient states. When applied to service life cycle model shown in Figure \ref{serviceSM} with initial state as Shutdown and goal state as Destroyed state, the algorithm produces 15 event sequences as shown in Figure \ref{serviceEvents}.

\subsection {Generating Permutation of Callback Sequences}

This section first describes how callback sequences are derived from the event sequences and then presents the logic to generate permutation of these callback sequences, and other callbacks (e.g., AUI callbacks).

\subsubsection{Deriving Callback Sequences}

Sections \ref{activityLCM}-\ref{serviceLCM} present activity and service life cycle models and the event sequences derived from these models. Each event in these sequences causes the system to invoke a specific sequence of callbacks in the respective component. Using Table \ref{activityTable}, Dexteroid obtains callback sequences for the 26 activity event sequences and then derives callback sequences using the life cycle callbacks defined in the activity code. It further removes duplicate callback sequences from the list. Similarly, Dexteroid obtains callback sequences for 15 event sequence of the service life cycle model. It removes duplicate event sequences which can produce the same callback sequence. For example, these two event sequences (createAndBind, start, unbind, stopAndDestroy; createAndBind, start, stop, unbindAndDestroy) produce the same callback sequence. It obtains 10 unique callback sequences from the 15 service event sequences shown in Figure \ref{serviceSM} and then derives callback sequences for analysis, given the actual service callbacks.

{\noindent \bf Example:} \label {scenariosExample} For the motivating example, Dexteroid identifies one activity and obtains the 26 callback sequences for the activity. With the given five activity callbacks, Dexteroid derives 12 unique callback sequences for the analysis.

\newcommand*{\permcomb}[4][0mu]{{{}^{#3}\mkern#1#2_{#4}}}
\newcommand*{\perm}[1][-3mu]{\permcomb[#1]{P}}

\subsubsection {Generating Permutation Sequences} 

The callback sequences are derived from event sequences for both activity and service life cycle models. They capture most of the callback orderings that can occur in Android applications. However, a  motivated attacker can still launch an attack using a specific ordering of callback-sequences, in place of individual callbacks. A callback sequence may be invoked more than once to launch the attack. Given the nature of a typical user interaction with an Android app, a user can invoke these callback sequences various times during the app life cycle. Thus, it is desired to analyze different orderings of callback sequences to detect targeted attacks hidden in the obscure program flows. Dexteroid uses permutation to obtain all possible orderings of these callback sequences. 

{\noindent \bf Permutation Unit:} After a createActivity event at the start in the activity event sequences in Figure \ref{activityEvents}, the remaining subsequences move activity from, and bring it back to its StaticPostResumed state. Since they can be repeated randomly in any order in this state, Dexteroid considers callback sequences derived from these event subsequences as permutation units. Similarly, the AUI-callbacks and miscellaneous callbacks can be invoked randomly in a StaticPostResumed state and are considered as permutation units. Dexteroid generates permutation of all permutation units and obtains the permutation callback sequences. It further prefixes these permutation callback sequences with a callback sequence from the createActivity event. This ensures that only feasible flows are generated during permutation (e.g., a createActivity event from a StaticPostResumed state would be an infeasible flow). The service does not have AUI callbacks but can have miscellaneous callbacks (e.g., {\ttfamily onLowMemory()}), Dexteroid generates permutation sequences for all permutation units.

\subsection {Detecting Malicious Behavior} \label{secMalBehavior}

Dexteroid can analyze all components iteratively in any order. For each component, it obtains permutation of callback sequences and analyzes them one by one. It performs static taint analysis on them to detect malicious behaviors (see Section \ref{taintAnalysisSec} for implementation of taint analysis). 

{\noindent \bf Threat Model:} Dexteroid currently produces warnings for two types of attacks: (1) leakage of sensitive information, and (2) sending SMS to premium-rate numbers. It tracks sensitive information produced by any of the pre-defined source APIs through callback sequences and reports a warning if such information is passed to any of the sink APIs. Additional warnings are reported if an app sends SMS to hard-coded number or automatically replies to incoming SMS messages. The warnings may reflect the legitimate functionality of the app, the determination of which is beyond scope of this work. Dexteroid does not detect attacks based on implicit flows \cite{implicitFlows}, Java reflection, native code, or dynamically loaded-libraries. It further does not consider attacks for detection that could be launched by exploiting system-level vulnerabilities \cite{dissectingAndroid, perils}.  

{\noindent \bf $m$-way Permutation:} A permutation of all permutation units can provide thorough analysis of the app code but it might be very resource-exhaustive task. Dexteroid instead considers $m$-way permutation of $N$ permutation units (i.e., $\perm{N}{m}$) where $m$ is $1\leq m\leq N$. An $m$-way permutation generates sequences by taking $m$ permutation units at a time out of $N$ permutation units. Dexteroid incrementally chooses $m$ values (starting from 1) for the analysis until it detects an attack. If an attack is found with lower $m$ values (e.g., $1$-way permutation), it can stop the analysis on the current app, and move to the next app. 

With $1$-way permutation, Dexteroid separately analyzes the 12 derived callback sequences and {\ttfamily onBtnClicked()} method of the motivating example but does not detect any attack. It then uses $2$-way permutation for the analysis and detects that many callback sequences can leak the deviceID. The $2$-way permutation detects the attack because the permutation sequences can contain {\ttfamily onUserLeaveHint()} method twice to transmit the deviceID to other variables. Specifically, all those callback sequences can detect the attack that contain ({\ttfamily onUserLeaveHint()}, {\ttfamily onUserLeaveHint()}, {\ttfamily onSaveInstanceState()}, {\ttfamily onRestoreInstanceState()}, {\ttfamily onResume()}) callbacks in the given order. This attack seems difficult to launch because it demands three event sequences to be triggered in a specific order but all of them are triggered often during a typical user interaction with the Android apps. 

\section{Implementation}

We have implemented a prototype of Dexteroid in Java. The tool works on Dalvik bytecode of Android applications. We chose Dalvik bytecode \cite{dalvikForm, DalvikByteCode} so that our analysis does not get affected by inaccuracies which are introduced during decompilation of an Android app into higher level languages \cite{BadAndroidDecompiler, Georgiev2012}. The following sections explain the major components which Dexteroid uses to detect malicious behaviors in Android apps.

\begin{algorithm}
\caption{Analysis Algorithm for an Activity}\label{activityAnalyzer}
\begin{algorithmic}[1]
\INPUT
\Statex $a$ \Comment Input activity for analysis
\Procedure{activityAnalyzer} {$a$}
	\State $S_{per} \leftarrow$ Get callback sequences from $m$-way permutation of $a$
	\ForAll {Sequence $P_{seq} \in S_{per} $}
		\ForAll {Method $d \in P_{seq}$}
			\State{\Call{methodAnalyzer}{$d$}}
		\EndFor		
	\EndFor	
\EndProcedure
\end{algorithmic}
\end{algorithm}

\subsection {Taint Analysis} \label{taintAnalysisSec}

We call system defined APIs which produce sensitive data as source APIs and the APIs which can potentially leak sensitive data out as sink APIs. A set of source and sink APIs are specified in a configuration file which Dexteroid uses during analysis. It marks a variable tainted if it obtains output of a source API. It performs forward taint flow analysis while keeping a track of source APIs and reports a warning if a sink API consumes data from a tainted variable. The reported warning contains specific sources that can potentially be leaked by a sink API. For each source, Dexteroid reports its data as well as its location in the code.

 \begin{algorithm}
\begin{algorithmic}[1]
\INPUT
\Statex $d$ \Comment Input method for analysis
\Statex $S_c$  \Comment Context stack
\Statex $S_m$  \Comment Method call stack
\Procedure{methodAnalyzer} {$d$, $S_c$, $S_m$}
	\State Remove back edges from CFG of $d$ and add new edges if needed
	\State $B_{rpo} \leftarrow$ Get basic blocks of $d$ in reverse post order\;
	\ForAll {BasicBlock $bb \in B_{rpo}$}
		\ForAll {Instruction $i \in bb$}
			\If{$i$ is a source API}
				\State Mark output as tainted
			\ElsIf {$i$ is a sink API}
				\State Report information leakage if an input variable is tainted
		      \ElsIf{$i$ invokes a method $d_{new}$ and $d_{new}$ is not on stack $S_m$ }
				\State Save current context of symbol space on context stack $S_c$
				\State Save $d_{new}$ information on call stack $S_m$
				\State Set relevant context for $d_{new}$
				\State returnValue $\leftarrow${\Call{methodAnalyzer}{$d_{new}$}}
				\State Pop call stack $S_m$
				\State Pop context stack $S_c$ and set output to current context
				\State Update symbol space with returnValue
			\ElsIf{$i$ is a return statement}
				\State Set return value for the caller method
			\Else
				\State Perform taint propagation on involved registers
			\EndIf
		\EndFor		
	\EndFor	
\EndProcedure
\end{algorithmic}
\caption{High-level Analysis Algorithm for a Method $d$}\label{methodAnalyzer}
\end{algorithm}

Dexteroid considers all components for analysis and handles them in an iterative manner. For a component such as activity, Dexteroid derives the callback sequences from the event sequences generated by $m$-way permutation. It then analyzes the callback sequences as shown in Algorithm \ref{activityAnalyzer}. For a callback sequence, it analyzes each callback in the sequence using Algorithm \ref{methodAnalyzer}. For each method $d$, Androguard \cite{Androguard} produces a control flow graph (CFG) consisting of basic blocks and directed edges between them. Dexteroid detects natural loops and back edges in the CFG using dominators \footnote{This loop detection technique does not work for irreducible CFGs which are rarely generated, however, in our day-to-day programming \cite{AhoCompilers}} \cite{AhoCompilers} and removes back edges from the CFG (line 2 in Algorithm \ref{methodAnalyzer}) . A back edge typically forms a loop, and goes from end of \emph{true} branch of the loop to the loop header. To combine results at merge point from both branches (paths) of the loop, Dexteroid adds new edge from end of \emph{true} branch to \emph{false} branch of the loop, as long as new edge does not introduce a new loop in the CFG. Dexteroid then analyzes each basic block in an order obtained by reverse-post-order (RPO) traversal of the CFG \cite{RPOTraversal}. During traversal, it maintains all analysis related information about predecessors and successors of a basic block which it later incorporates to perform path-insensitive analysis. It combines tainted data from different paths (predecessors) at merge points in the CFG \cite{staticAnalysis}. This path-insensitive approach is more practical and less expensive as compared to path-sensitive analysis, though it may produce false positives or false negatives in some cases.

\subsubsection{Context-Sensitive Analysis} \label{secContextSensitive}

Dexteroid analyzes each (callback) method $d$ using Algorithm \ref{methodAnalyzer}. It iterates over instructions of each basic block where an instruction contains one of the 246 Dalvik opcodes \cite{DalvikByteCode}. Dexteroid implements opcode-specific parsers and handlers which perform analysis on respective instructions. During the analysis, Dexteroid maintains a symbol space (consisting of symbol tables at multiple levels) to keep up-to-date information about variables (registers) such that an instruction gets access to only those variables which have been defined at current basic block, method, or class level, and at global level (e.g., static fields). 

Dalvik bytecode provides five {\ttfamily invoke-}\emph{kind} instructions (e.g., {\ttfamily invoke-virtual}) to invoke all types of APIs and methods \cite{DalvikByteCode}. If an instruction invokes a developer-defined method or a library method for which definition exists in the app code (line 10 in Algorithm \ref{methodAnalyzer}), its opcode-specific handler saves current context of symbol space on context stack, finds the callee method by its signature, sets up context for the callee method and invokes analysis on it. After completing analysis on the callee method, the analysis execution automatically returns back to the caller-instruction handler which retrieves old context from the context stack and continues its analysis (line 10-17). Dexteroid obtains \emph{call-site context sensitivity} by maintaining context automatically through the instruction-specific handlers. 

{\noindent \emph{Handling recursive-functions}:} In addition to context stack, Dexteroid also maintains a function or method call stack which it consults with before making the context-switch for a callee method. If a method call with a matching signature does not exist in the stack, it adds its entry into the stack and makes context switch otherwise it ignores the method call. After a callee method returns, its entry from the stack is removed and analysis of the caller method continues. Such an approach may not be completely sound to analyze recursive functions but this does help Dexteroid avoid winding up in an infinite loop, especially in two cases: (1) direct recursion in which a method calls itself to implement the desired functionality, and (2) indirect recursion in which multiple methods call each other to form recursion (e.g., method $a$ calls method $b$ and method $b$ calls method $a$). 

\subsubsection{Object and Field Sensitivity} \label{objSensitivity}

To achieve high level accuracy, one important problem is how to obtain object-level sensitivity to resolve aliasing during the analysis. In Dalvik bytecode, objects are manipulated using variables (registers). Corresponding to each such variable, Dexteroid maintains one unique entry in the symbol tables. Each entry instance consists of a \emph{name} field and a \emph{details} field of type `EntryDetails'. To resolve aliasing issues, Dexteroid obtains shallow copy of a symbol table entry such that all changes made later to shallow-copied object are automatically reflected in all aliases of that object. A shallow-copied entry's \emph{details} object keeps pointing to the same \emph{details} object of original entry while a deep-copied entry's \emph{details} object gets a new copy of original \emph{details} object at a different memory location. During analysis, when one shallow-copied entry is assigned a new object, other aliases still keep pointing to the original entry in the memory. For primitive data types and immutable objects (e.g., {\ttfamily String}), Dexteroid considers deep copy of symbol table entries for manipulations. This kind of heap abstraction ensures that all aliases are always pointing to the same object. Thus, when a new method is called and context-switch happens, Dexteroid sets up formal parameters by making shallow copy and deep copy of the actual parameters (depending upon their data types). Similarly, Dexteroid obtains field sensitivity by storing each field within `fieldList' of each symbol table entry's \emph{details} object. Each field in the `fieldList' is itself a symbol table entry, and recursively has its own `fieldList'. Dalvik bytecode \cite{DalvikByteCode} provides a set of {\ttfamily iget} and {\ttfamily iput} instructions to access individual fields of an object. Using these instructions, Dexteroid theoretically can maintain fields of any depth during the analysis. This helps in performing very precise taint analysis for user-defined classes with different fields (e.g., linked list). Consider an example in which a linked list of objects of a user-defined class 'Node' (with 'next' and 'value' fields) is defined,  and a specific node of the linked list (e.g., \emph{head.next.next.value}) is set to tainted. Dalvik bytecode uses {\ttfamily iput} and {\ttfamily iget} instructions to set and get these fields while to perform analysis, instruction-specific handlers of Dexteroid recursively obtain, and store \emph{next} and \emph{value} fields in 'fieldList' of the \emph{head} entry. Thus, when a specific node is accessed, Dexteroid obtains the node and reports a warning if its \emph{value} field is tainted.


{\noindent \bf Example:} Listing \ref{listing:objSenstvty} shows an object-sensitivity example, derived from example given in \cite{FlowDroid}. The code creates {\ttfamily x1} and {\ttfamily y1} instances (line 2-3) for which Dexteroid creates two symbol table entries and stores {\ttfamily x1} symbol table entry into `fieldList' of {\ttfamily y1}'s \emph{details} object. The variable {\ttfamily x2} gets reference of instance {\ttfamily x1} from {\ttfamily y1} at line 4. Correspondingly, Dexteroid makes shallow copy of the original symbol table entry \emph{pointing to} {\ttfamily x1} and renames it for {\ttfamily x2}. At line 5, Dexteroid invokes {\ttfamily foo()} and passes a copy of {\ttfamily y1} symbol table entry as {\ttfamily yP}. Thus, when {\ttfamily xT.val} is assigned a source value (line 11), Dexteroid taints its corresponding symbol table entry object in the memory. Upon return from analysis of {\ttfamily foo()}, changes made to {\ttfamily x1} entry are automatically reflected in {\ttfamily x2} entry.  Line 6 passes tainted value obtained from {\ttfamily x2.getVal()} to the sink API for which Dexteroid reports an information leakage warning.

\definecolor{mygreen}{rgb}{0,0.6,0}
\definecolor{mygray}{rgb}{0.5,0.5,0.5}
\definecolor{mymauve}{rgb}{0.58,0,0.82}

\lstset{language=Java, basicstyle=\scriptsize\ttfamily, numbers=left, stepnumber=1, keywordstyle=\color{blue},    
frame=single, numbersep=5pt, showspaces=false, numberstyle=\scriptsize\color{black}, stringstyle=\color{mymauve},
showstringspaces=false, showtabs=false, tabsize=2,
 breaklines=true, breakatwhitespace=false, caption={Object-Sensitivity Example} , label={listing:objSenstvty}}

\noindent\begin{minipage}{7cm}  
\begin{lstlisting}
void main(){
	x1 = new X(); 
	y1 = new Y(x1);
	x2 = y1.getX();
	foo(y1);
	sink(x2.getVal());
}
void foo(Y yP){
	xT = yP.getX();
	w = source();
	xT.setVal(w);
}
\end{lstlisting}
\end{minipage} 

{\noindent \emph{Taint preservation along paths of a CFG:}} Dexteroid analyzes basic blocks of a CFG in a fixed RPO order and conservatively combines results at merge points of the CFG. It maintains an {\ttfamily OUT} symbol table for each basic block after completing its analysis. An {\ttfamily OUT} symbol table reflects the method state after analysis of a basic block and is used as an input by its successor basic blocks. At a merge-point basic block of a CFG, results are combined (conservatively) from {\ttfamily OUT} symbol tables of all its immediate predecessors. This means that if a tainted variable gets untainted along one path of CFG but propagates taint on the other path, it will contain tainted information at the merge point. However, since Dexteroid maintains shallow-copied entries corresponding to heap objects, the first analyzed path may untaint a tainted object and cause the analysis to modify, and untaint its corresponding symbol table entry in the memory. When the second path is analyzed, it gets shallow-copy of the entry which would be untainted now (caused by analysis of the first path). This will cause losing taints at the merge point. To address this issue, Dexteroid sets a separate \emph{deep-copied} duplicate of whole {\ttfamily OUT} symbol table ({\ttfamily OUT$_d$}) for each basic block after its analysis. Thus, if an entry is modified and untainted along one path, the entries in {\ttfamily OUT$_d$} are not modified, and still contain taints for propagation along the other paths. An entry on any path can now get preserved taints from the {\ttfamily OUT$_d$} and propagate them to the merge point. The above discussion assumed only one merge point in the CFG but in practice, there might be many merge points for different predicates or loops. Another issue may arise during the analysis in which an object can point-to different objects along different paths of a CFG. A path-sensitive analysis can help Dexteroid to analyze separate program paths, and resolve this issue but such an analysis is known to be a very expensive analysis. 

\subsubsection{Flow-Sensitivity} 

As shown in Algorithm \ref{methodAnalyzer}, Dexteroid sequentially analyzes instructions in a basic block while maintaining all up-to-date information about the variables. It performs on-demand updates on the variables and hence automatically maintains flow-sensitivity during the analysis.

{\noindent \bf Example:} Listing \ref{listing:flowSensitivity} shows an example (derived from an example in \cite{FlowDroid}) where flow-insensitive analysis would have reported two information leakage warnings at line 4 and line 6. At line 4, {\ttfamily x2.getVal()} is not tainted yet, Dexteroid does not report anything but {\ttfamily x1} gets tainted at line 5, Dexteroid then correctly reports information leakage warning at line 6.

\definecolor{mygreen}{rgb}{0,0.6,0}
\definecolor{mygray}{rgb}{0.5,0.5,0.5}
\definecolor{mymauve}{rgb}{0.58,0,0.82}

\lstset{language=Java, basicstyle=\scriptsize\ttfamily, numbers=left, stepnumber=1, keywordstyle=\color{blue},    
frame=single, numbersep=5pt, showspaces=false, numberstyle=\scriptsize\color{black}, stringstyle=\color{mymauve},
showstringspaces=false, showtabs=false, tabsize=2,
 breaklines=true, breakatwhitespace=false, caption={Flow-Sensitivity Example} , label={listing:flowSensitivity}}

\noindent\begin{minipage}{7cm}  
\begin{lstlisting}
void main(){
	x1 = new X();
	x2 = x1;
	sink(x2.getVal());
	x1.setVal(source());
	sink(x2.getVal());
}
\end{lstlisting}
\end{minipage} 

\subsection {Handling APIs} \label{apiHandlingSection}

While source and sink APIs are handled to produce and consume sensitive data respectively, other set of APIs may play their part in taint propagation. These APIs belong to different libraries such as Android, Java, and Apache etc. We have implemented specific handlers for most frequently invoked APIs, extracted from our evaluation dataset. For example, call to an API such as  {\ttfamily Ljava/lang/StringBuilder;->append} is delegated to its specific handler which makes relevant variables tainted or untainted based on information obtained from the symbol tables and API documentation. Similarly, some native APIs such as {\ttfamily Ljava/lang/System;->arraycopy} have specific handlers for performing analysis on them. For remaining large set of APIs, Dexteroid has default {\ttfamily invoke} handlers which mark a caller object or returned value as tainted if any of its input parameters or the caller object is tainted. However, the analysis by default handlers is potentially unsound and may produce false positives because output from an object containing sensitive information may not always be sensitive. For example, for an {\ttfamily ArrayList} API ({\ttfamily Ljava/util/ArrayList;->isEmpty()Z}), even if the list contains sensitive data, the returned boolean value from the list may not provide any useful or sensitive information in most of the cases. 

Dexteroid defines default handlers for all kinds of {\ttfamily invoke} instructions \cite{DalvikByteCode}. If definition for the instruction-invoked method or API does not exist in the code, one of the default {\ttfamily invoke-}\emph{kind} handlers (e.g., InvokeStaticDefaultHandler) performs taint analysis on it. For example, {\ttfamily invoke-static} makes calls to static methods and APIs and does not need any caller or receiver object to invoke a method or API but {\ttfamily invoke-virtual} typically requires it. When analyzing an API or a method call, the {\ttfamily invoke}-specific handlers mark output value or caller object as tainted or not-tainted based on input parameters and the caller object, if any. For collection objects (e.g., array, hashtable), Dexteroid considers the object as tainted if any of its elements gets tainted. The object stays tainted even if the tainted element (specific indexed value) in the object becomes untainted. However, it becomes untainted when a \emph{new} untainted object is assigned to it. For collection objects, it is difficult to statically get exact index values, so marking the whole object as tainted ensures that we do not miss any attack, though it may generate false alarms in some cases.

\subsection {Handling Static Flow Discontinuity} \label{handlingAPIs}

Dexteroid handles static flow discontinuity issue \cite{staticDiscontinuity} during the analysis which arises when a component or method is invoked using a prototype which does not match with its original prototype definition. For example, {\ttfamily run()} method of a thread is executed when a call to its {\ttfamily start()} method is made. Android dynamically resolves such issues but a static analysis tool must handle such discontinuity issues during the analysis. Similarly, an {\ttfamily AsyncTask} \cite{asynchTask} is started by making a call to {\ttfamily execute()} method but such a method is not defined for a task. In fact, following its life cycle, Android invokes its four callbacks ({\ttfamily onPreExecute()}, {\ttfamily doInBackground()}, {\ttfamily onProgressUpdate()}, and {\ttfamily onPostExecute()}) in the given order. Dexteroid handles both component and method discontinuities by precisely making context switch during the analysis. During preprocessing phase before the analysis, Dexteroid determines type of a class (e.g., task, thread) based on its parent class or defined (callback) methods. Using this information, it determines if a caller to an {\ttfamily execute()} method is an {\ttfamily AsyncTask}, and if so, it delegates analysis to the task-specific handler by making a context-switch (as described in Section \ref{secContextSensitive}). The task handler analyzes all callbacks following its life cycle model before returning back to the task-executor (or caller) method. Similarly, a call to {\ttfamily start()} method is delegated to a thread handler which analyzes its {\ttfamily run()} method before returning back to the caller method.

\subsection{Comparison with Other Tools} \label{secComparison}

This section compares the analysis approach of Dexteroid with those of other static analysis tools such as FlowDroid \cite{FlowDroid} and DroidSafe \cite{DroidSafe}, as below.

\begin{itemize}

\item \emph{Analysis-methodologies for event-triggered callbacks:} Dexteroid analyzes permutation of callback sequences which are systematically derived from the event sequences, and are guaranteed to be valid callback sequences. In comparison, FlowDroid builds a CFG of the callbacks (i.e., a \emph{dummy main} method) and performs analysis on paths of the CFG. This CFG-based approach may not accurately model and analyze Android applications because they are event-driven programs by nature. Events can invoke specific callback sequences in the program and can themselves be triggered multiple times during the program execution. The CFG of callbacks, however, lacks guard conditions and does not reflect conditional flows between the callbacks. A path-insensitive approach (such as FlowDroid) combines results at merge points of the CFG which affects both accuracy as well as precision. Thus, to bypass such an analysis, an attacker can design attacks that are launched by specific ordering of event sequences, such as given in the motivating example. In comparison, DroidSafe \cite{DroidSafe} considers all possible orderings of callbacks for the analysis. The all possible ordering of callbacks violates the activity life cycle model and can cause false positives in the analysis results \cite{DroidSafe}. 
 
\item \emph{Techniques for maintaining flow-sensitivity and call-site context-sensitivity:} Dexteroid is inherently flow-sensitive and context-sensitive which leads to higher accuracy and higher precision in the analysis results. It performs taint analysis by analyzing callbacks guided by the callback sequences and within a (callback) method, it analyzes all instructions in a sequential order, maintaining flow-sensitivity. Similarly, call-site context-sensitivity is ensured by an on-demand analysis of the invoked method, as described in Section \ref{secContextSensitive}. In comparison, FlowDroid uses \emph{context-injection} in forward-analysis and backward-analysis to maintain context-sensitivity. Furthermore, it relies on \emph{activation-statements} to achieve flow-sensitivity. Such an approach might be error-prone if all cases are not handled carefully (We found in our experiments that FlowDroid is not able to maintain flow-sensitivity for globally-defined objects, and can cause false positives. In one case, for a globally-defined object {\ttfamily o1}, a call to method {\ttfamily m1()} initializes {\ttfamily o1}, sets {\ttfamily o1.f=deviceID}, and then sets {\ttfamily o1.f=""}. A call to method {\ttfamily m2()} after {\ttfamily m1()} invokes {\ttfamily o1.p()} which gets {\ttfamily f} and passes it to a sink API. FlowDroid reports an information leakage warning for this. This false positive arises probably because of an on-demand backward analysis which is invoked when a heap object is tainted. The backward analysis later causes forward analysis to map back to the caller of {\ttfamily m1()} from where it then spawns a forward analysis. This forward-analysis analyzes {\ttfamily m2()} which then leads to reporting of information leakage (i.e., a false positive)). Analysis of DroidSafe \cite{DroidSafe} is context-sensitive but flow-insensitive. Its context-sensitivity is specific to object-sensitivity: for a given method, it can accurately compute the receiver heap object on which the method is called. However, it is unclear from the paper that how the call-site context-sensitivity is maintained for static and direct method calls.

\item \emph{Alias-analysis approaches:} Alias-analysis determines if two pointer variables or references can point-to the same memory location during the program execution. Corresponding to heap objects, Dexteroid maintains symbol table entries in the symbol tables. Dexteroid resolves aliasing through shallow-copied entries such that both entries point-to the same memory location (see Section \ref{objSensitivity}). Whenever a heap object is tainted, FlowDroid spawns backward-analysis to taint its aliases which can further spawn new forward and backward analyses and so on. It uses access-paths of specific length to track objects and their fields, and propagates access-paths during the analyses. Our experiments show that alias-analysis for global-level objects also gets affected by backward and forward analyses, as discussed above. DroidSafe, in comparison, is very accurate in resolving-aliases for object-sensitive analysis. It maintains \emph{k-}level deep object-sensitivity for the analysis.

\item \emph{API-handling techniques:} In addition to source and sink APIs, other APIs can be involved in taint propagation. Dexteroid implements specific handlers and default handlers to analyze such APIs  (see Section \ref{handlingAPIs} for details). Similarly, FlowDroid has a concept of \emph{taint-wrappers} to model and analyze these APIs. However, both Dexteroid and FlowDroid do not analyze the internal code of these APIs and may cause inaccuracy during the analysis. In comparison, DroidSafe's accurate analysis stubs provide highly accurate execution model of Android for a static analysis tool. Furthermore, it maintains deep object-sensitivity for those API calls to perform highly accurate analysis of the program.

\end{itemize}

\section{Experimental Evaluation}

To evaluate the effectiveness of Dexteroid, we perform a series of experiments to detect two types of attacks: (1) information leakage attack, and (2) SMS-sending attack. An information leakage attack leaks sensitive or private information and in an SMS-sending attack, an application can send SMS to premium-rate numbers.

\subsection{Detection of Information Leakage}

We evaluate Dexteroid against real-world Android applications from Google Play (GPlay) and Genome Malware (Genome) \cite{dissectingAndroid}, and further validate its effectiveness against a benchmark suite of Android apps called DroidBench \cite{FlowDroid}. In addition, we compare our results, where applicable, with previous open-source, state-of-the-art static analysis tool called FlowDroid \cite{FlowDroid}. FlowDroid becomes an ideal candidate for comparison because like Dexteroid, it performs life-cycle aware, context, object, flow, and field-sensitive taint analysis to detect privacy leaks. Furthermore, both tools do not handle inter-component and inter-app communication for the analysis. We use FlowDroid's source code version \cite{FDWiki} obtained on September 9, 2014 in all our experiments, unless specified otherwise. All experiments are run with default options set by FlowDroid, using {\ttfamily -Xmx12g} with no performance-improving arguments. 

\subsubsection{Evaluation on Real-World Android Apps} \label{evalReadWorld}

{\noindent \bf Experimental Setup:} We run both tools on top free 1526 GPlay apps and 1259 Genome apps \cite{dissectingAndroid}. Each experiment for both tools is run with Dexteroid-provided source and sink API set. Both tools derive these API sets from Susi project \cite{Susi}. However, to avoid missing any privacy leak, FlowDroid uses over-approximation in selection of source and sink APIs which may cause many false warnings. For example, it assumes output from APIs related to different collection objects as source APIs (e.g., {\ttfamily bundle.getDouble()}), even if the object contains non-sensitive data. Similarly, a flow to an API such as {\ttfamily intent.setClassName()} is reported as information leakage. Dexteroid-provided source and sink API set (68 sources and 75 sinks) does not contain such APIs. Moreover, under time constraints, all our experiments with Dexteroid are run with $1$-way permutation, unless specified otherwise. However, it might be interesting to analyze applications with all-way permutation (i.e., considering $m$=$N$ in $\perm{N}{m}$). An analysis time for an app is heuristically chosen to be 10 minutes.

{\noindent \bf Evaluation Criteria:} Android apps from Google Play (GPlay) are assumed to be benign apps with no defined set of true information leakage warnings. Similarly, Genome samples are known malware and have reported malicious behaviors \cite{dissectingAndroid} but to the best of our knowledge, the number of unique information leakage warnings produced by each sample are not known. Thus, to evaluate analysis results, we first formally define the number of warnings for an app set. We denote a set of maximum number of unique true warnings an app set can theoretically have with $W_{T}$, a set of combined true warnings detected by existing tools (such as Dexteroid and FlowDroid) with $W_{CO}$, and a set of unique true warnings produced by Dexteroid and FlowDroid with $W_{DT}$ and $W_{FD}$, respectively. 
    
\begin{align}
W_{CO} & =  W_{DT} \cup W_{FD}\label{eq1}\\ 
W_{CO} & \subseteq  W_{T} \label{eq2}
\end{align}

We use $W_{CO}$ to evaluate analysis results (e.g., recall calculations) of both tools and consider it as a reasonable assumption, given the fact that no set of true warnings are available for existing apps. Hopefully, as true warning sets of other tools are added, the $W_{CO}$ approaches $W_{T}$. A unique warning for an app must meet two criteria: (1) it has a unique combination of a set of source APIs and a sink API, (2) for another warning with the same sink API, its source API set is not subsumed by that of the other. In our evaluation, we consider only unique warnings produced by both tools.

\begin{figure*}[ht]
\centering
\includegraphics[width=120mm]{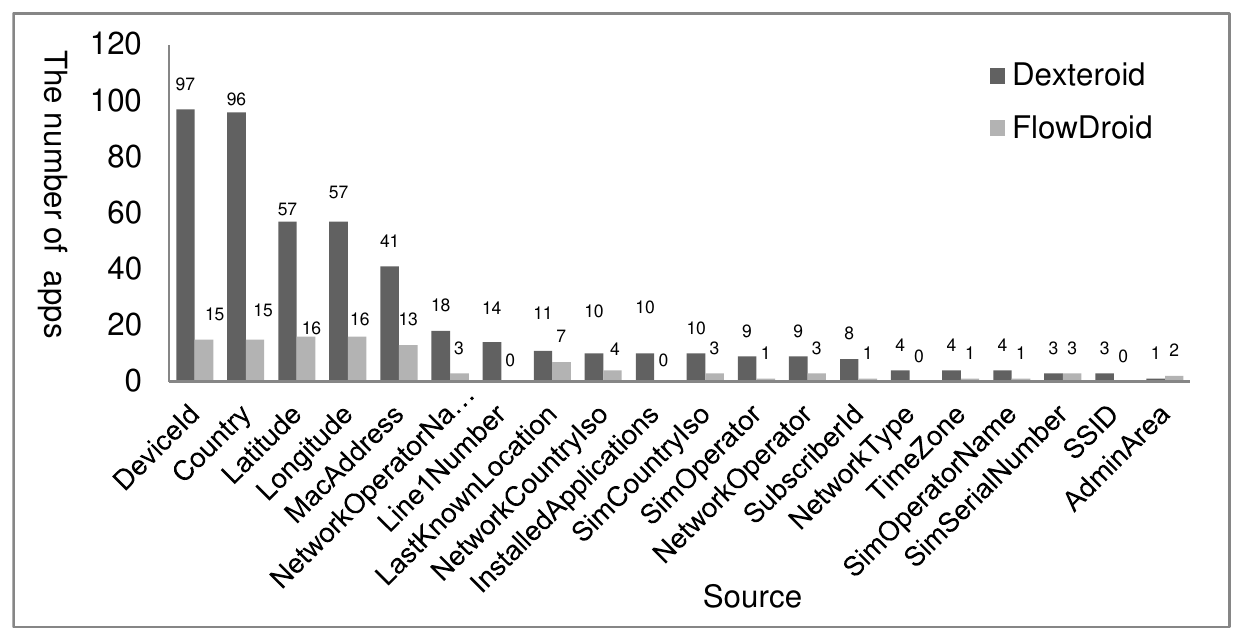}
\caption{Information Leakage Detection by Dexteroid and FlowDroid in 1526 Google Play Apps}
\label{gplaySrcSink}
\end{figure*}

{\noindent \bf Evaluation on Google Play Apps:} When applied on GPlay apps, both tools report many applications leaking sensitive information as shown in Figure \ref{gplaySrcSink}. The x-axis shows the top 20 source API data which are leaked and y-axis shows the number of apps which leak the specific sensitive data (obtained from a source API). For example, Dexteroid reports that `deviceID' and `country' information can be leaked by 97 and 96 apps, respectively. Similarly, 14 apps are reported to leak phone number obtained using {\ttfamily getLine1Number()} API.

For 1526 GPlay apps, Dexteroid produced 661 warnings and FlowDroid produced 201 information leakage warnings. The manual verification of such a large number of warnings is a very time consuming task. It took us one hour on average to verify a warning produced by either tool because a warning may use many (e.g., more than 20) deep level method calls to obtain, and leak sensitive information. The manual verification becomes more difficult by obfuscation techniques used by developers to save their code from stealing. Thus, to evaluate our approach in a meaningful manner, we perform sampling at applications level and then apply both tools on the same set of sampled apps. A random sampling is performed to pick 10\% of the apps (152 apps). Table \ref{expResults} shows experimental setup, and evaluation results for both tools. In experiment\#1, Dexteroid produces 68 unique information leakage warnings; 8 of them are false positives. The false alarms are raised because of the default handlers as well as marking collection objects (e.g., {\ttfamily HashMap}) as tainted, even when value at a specific index is tainted only (see Section \ref{apiHandlingSection}). FlowDroid produces 16 warnings for the 152 apps which however include 5 false positives. Dexteroid detects all the remaining 11 warnings produced by FlowDroid and overall achieves 100\% recall and 88.2\% precision. Its recall is 100\% because (1) all experiments are run independently with their specific configurations, (2) in experiment\#1, Dexteroid detects all the true warnings detected by other tools (i.e., FlowDroid in this case) (see Section \ref{evalReadWorld}). In comparison, FlowDroid achieves 18.3\% recall and 68.7\% precision.

\begin{table*}[ht] 
\caption{Evaluation of Dexteroid and FlowDroid on Sampled Google Play Apps}
\centering 
\scriptsize
\begin{tabular}{l | l | m{.5cm}  m{.5cm} | m{.5cm}  m{0.5cm}  m{.9cm} | m{.5cm}  m{0.6cm}  m{.9cm}}
\hline
&{\multirow{2}{*}{Item}} & \multicolumn{2}{c|}{Experiment\#1} & \multicolumn{3}{c|}{Experiment\#2}  & \multicolumn{3}{c}{Experiment\#3}\\ \cline{3-10} 
    &   & DT & FD &  DT & FD & FD-RE &  DT & FD' & FD'-RE \\  
\hline
\parbox[t]{0.1mm}{\multirow{4}{*}{\rotatebox[origin=c]{90}{Setup}}} & \#Test Apps & 152 & 152 & 158 & 158 & 158 & 158 & 158 & 158\\   
& $m$-way Permutation & $1$ & N/A &  $1$ & N/A & N/A & $1$ & N/A & N/A \\   
&Max Time (min) &  10 & 10 & 30 & 30  & 30  & 30 & 30  & 30\\   
&Source \& Sink API Set &  DT & DT & DT & DT & DT  & DT & DT & DT\\   
&Warning Type & I & I  & I & I & I  & I & I & I \\
\hline
\parbox[t]{2mm}{\multirow{8}{*}{\rotatebox[origin=c]{90}{Results}}} &\#Reported Warnings & 68 & 16 & 155 & 26 & 27 & 155 & 140 & 145 \\  
&\#True Warnings & 60 & 11 & 138 & 9 & 8 & 138 & 85 & 89 \\   
&\#Combined Warnings ($W_{CO}$) & 60 & 60 & 138 & 138 & 138 & 157 & 157 & 157\\  
&Recall(\%) & 100 & 18.3 & 100 &6.52 & 5.8 & 87.8 & 54.14 & 56.68 \\   
&Precision(\%) & 88.2 & 68.7 & 89.03 & 34.6 & 30.76 & 89.03 & 60.7 & 61.3 \\   
&F-1 Score & 0.93 & 0.29 & 0.94  & 0.11 & 0.1 & 0.88 & 0.57  & 0.59  \\   
&\#Killed(\%) & 24  &39  & 0 & 0 & 3.16 & 0 & 0 & 1.89 \\   
&\#Finished(\%) & 76 & 61 & 100 & 100 & 96.84 & 100 & 100 & 98.11 \\  
\hline
\multicolumn{10}{l}{
  \begin{minipage}{11cm}
    \scriptsize DT=Dexteroid, FD=FlowDroid,  FD-RE= FlowDroid with flows from reverse-engineered life cycle models, FD'=FlowDroid's latest source code, FD'-RE= FD' with flows from reverse-engineered life cycle models,  I=Information leakage
  \end{minipage}
}\\
\end{tabular} 
\label{expResults} 
\end{table*} 

FlowDroid reports 11 out of 60 true warnings in experiment\#1. This low-detection could be due to FlowDroid's inaccurate modeling of Android components, or its incomplete taint tracking process for the analysis. The analysis results of both tools can be further affected by limited time or memory for the experiment \cite{FDMailing}(e.g., FlowDroid's analysis is not completed under 10 minutes for 39\% of the apps). To address these issues, we first randomly pick 158 apps based on the experiment\#1 such that both tools do not run out of time or memory during the analysis. We then modify FlowDroid's source code to include additional callback flows from our reverse-engineered activity and service life cycle models. We add seven flows for an activity (e.g., from {\ttfamily onStart()} to {\ttfamily onStop()}) and three flows for a service (e.g., from {\ttfamily onStartCommand()} to {\ttfamily onUnbind()}) in the modified code (FD-RE for short). Experiment\#2 is run on 158 apps and evaluation results are shown in Table \ref{expResults}. Dexteroid reports 155 warnings while FlowDroid reports 26 warnings. Dexteroid reports false alarms because of default handlers and false positives for FlowDroid arise mostly due to its over-approximations for intent-communication (e.g., considering an incoming intent as a source). FD-RE reports one additional warning which is false positive but interestingly, it misses also one true warning reported by FlowDroid. The missed warning is reported by {\ttfamily com.fingersoft.nightvisioncamera} app for which FD-RE is not able to complete its analysis under 30 minutes due to additional callback flows for activity and service components.

FlowDroid has been under active development since its release as an open source tool\footnote{FlowDroid's Android develop branch alone has 567 commits on its Github account at the time of publication.}. Thus, to examine how Dexteroid's prototype implementation would perform against FlowDroid's updated version, we obtain FlowDroid's latest source code (FD' for short) \cite{FDWiki} on August 28, 2015, and add flows from our component life cycle models in the modified FD' (FD'-RE for short). For high precision, none of the performance-improving options is used for FD' and FD'-RE. Experiment\#3 is run on 158 apps of experiment\#2 and evaluation results are shown in Table \ref{expResults}. FD' produces 140 warnings of which 55 warnings are false positives. FD'-RE reports five additional warnings of which one is false positive and four are true positives. Even with additional callback flows, FD'-RE's recall (57\%) remains low than Dexteroid's recall (88\%). This could be due to its dummy-main method based approach or incomplete taint tracking during the analysis or both (see Section \ref{secComparison}). Dexteroid also misses some of the true warnings reported by FD' and FD'-RE. They are missed because Dexteroid's prototype implementation lacks complete modeling of different Android APIs and components. With future improvements, we hope to detect all the missed warnings reported by FD' and FD'-RE. In addition, we are interested to perform more experiments in the future in which we could run both tools (Dexteroid, and FlowDroid's latest code with additional callback flows) on all 1526 GPlay apps and 1259 Genome apps for at least 10 hours per app with a reasonable amount of memory. In our earlier experiments with 1526 GPlay apps, FlowDroid's analysis was killed after ten minutes for 42\% of the apps (see Section \ref{performanceEvaluation}). Similarly, in some cases, FlowDroid may need more than 500 GB of memory to complete its analysis \cite{FDMailing}. Thus, ideally we would like to run our new experiments with enough amount of time and memory so that both tools can complete their analyses for all the apps. These experiments may take a few months while FlowDroid is continuously being updated in parallel; we will pick one specific version of FlowDroid (such as FD'-RE) and run experiments with it.

{\noindent \emph{Phone-Number Leaking Apps}:} The experiments in Table \ref{expResults} report information leakage warnings for all type of source APIs. Some APIs produce more sensitive data (e.g., phone number) than others (e.g., country). The leakage of phone number may expose users to receive unwanted (e.g., advertisement) phone calls, and spam messages. To detect such phone number leakage attacks, both tools are run on 1526 GPlay apps. Dexteroid reports that 14 apps can potentially leak user's phone number as shown in Table \ref{phoneNoLeakingAppsTab} while FlowDroid does not report any warning. Most of the apps send phone number using {\ttfamily http} connection or SMS while 2 apps (\#1 and \#2) write phone number into a "device\_info.txt" file on external memory card. Two apps (\#3 and \#4), apparently from the same developer, contain same (non-advertisement) code and produce the same four warnings each. They start threads from four different activities ({\ttfamily FirstRun}, {\ttfamily Login}, {\ttfamily PaidReturn}, {\ttfamily Subscription}), obtain information such as deviceID, phoneNo and subscriberID and send it to a hard-coded URL starting with {\url{http://www.surveynotify.com}}. The {\ttfamily FirstRun} activity is started when user opens the app for the first time and the activity leaks the sensitive information in plain text. Similarly, app\#5 obtains phoneNo in the {\ttfamily onStartCommand()} method of its {\ttfamily SimCardCheck} service and leaks this information by sending a text message (SMS) in {\ttfamily sendSMS()} method of a {\ttfamily GF} class.

All 14 apps produce 20 warnings: two of them are false positives. Dexteroid reports that the app\#14 in Table \ref{phoneNoLeakingAppsTab} leaks phone number and device information at two places but our investigation reveals that they are false positives. The string contains phone number and other device information but later device information is extracted and passed to the sink APIs. Dexteroid's default handlers pass tainted data from a string (operation) as it is and hence lead to producing the false positives. The customized handlers or stubs for specific APIs (e.g., {\ttfamily String.split()}) may help in eliminating such false alarms.

\begin{table}[ht] 
\caption{Detection of Phone-Number Leaking Apps By Dexteroid in Google Play Apps}
\centering 
\scriptsize
\begin{tabular}{l  l l c c } 
\hline
\# & App ID& Version & True Positive & False Positive\\  
\hline 

1 & cz.aponia.bor3 & 3.10.16 &  \checkmark  & \\
 2 &cz.aponia.bor3.truck & 3.10.28987 &  \checkmark  & \\
  3 & thecouponsapp.coupon & 9.52 &  \checkmark  & \\ %
4 & gas.coupons & 9.15 &  \checkmark  & \\
 5& com.alienmanfc6.wheresmyandroid & 5.2.1 & \checkmark & \\
 6 &com.fatsecret.android & 3.1.1 &  \checkmark  & \\
 7 &com.meraki.sm & 0.9.51 &  \checkmark  & \\
 8 &com.tma.frontflip & 5.0.37 &  \checkmark  & \\
 9 & ipnossoft.rma.free & 2.3.3 &  \checkmark  & \\
 10 &com.facetime\_plus.trendy & 1.0.1 &  \checkmark  & \\     
 11 & kst.DailyTextLite2 & 1.3.3 &  \checkmark  & \\     
 12 & com.gau.go.launcherex & 5.02.1 &  \checkmark  & \\
 13 & com.yinzcam.nfl.seahawks & 1.1.1 &  \checkmark  & \\
 14 & com.unionbank.ecommerce.mobile.android & 2.10.0.2 &   &\checkmark \\ 
\hline 
\end{tabular} 
\label{phoneNoLeakingAppsTab} 
\end{table}

{\noindent \emph{Supplementary Callbacks}:} The CSS model of activity life cycle includes six supplementary callbacks in addition to seven callbacks shown in Figure \ref{coreLifeCycle}. With 1526 GPlay apps, Dexteroid reports one information leakage warning which is triggered from a supplementary callback method. One app {\ttfamily com.pelmorex.WeatherEyeAndroid} ("The Weather Network") starts an asynchronous task from a supplementary callback {\ttfamily onRestoreInstanceState()}. The task obtains geo-location coordinates, and deviceID and makes a request to a third party server using a sink API. Dexteroid passes parameters to the asynchronous task and performs context-sensitive analysis to successfully detect this information flow.

{\noindent \emph{$2$-way Permutation}:} To find out real-world Android apps that use event permutation to leak information, we run Dexteroid on 1526 GPlay apps with $2$-way permutation and maximum allowed time per app to 30 minutes. We did not find any app which exhibits specific malicious behavior using event permutation. However, we find one such information flow in a `Virtual Zippo Lighter' app which uses combination of a miscellaneous callback method ({\ttfamily onGeocodeTaskComplete()}), and an AUI callback method ({\ttfamily onEditorAction()}) to pass sensitive information to a sink API. With this information flow, the app passes geo-location address to obtain geo-location coordinates from Google servers. Dexteroid detects only one such information flow using $2$-way permutation. However, it might be interesting to see evaluation results with higher order permutation orders (e.g., all-way).

\begin{figure*}[ht]
\centering
\includegraphics[width=120mm]{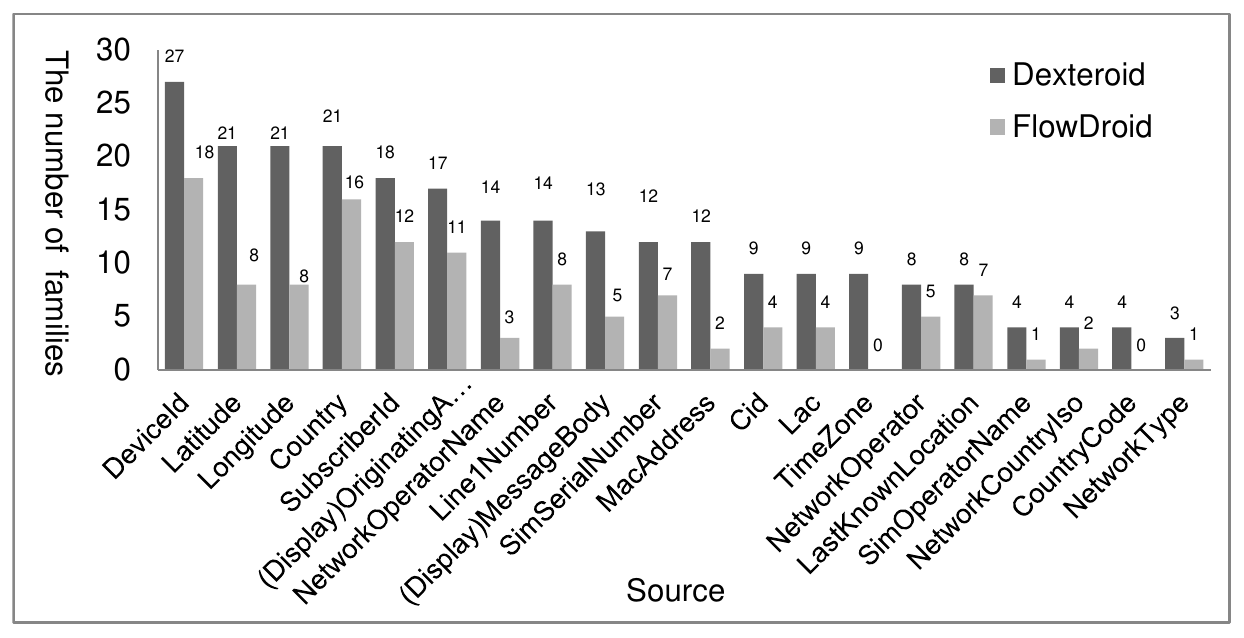}
\caption{Information Leakage Detection by Dexteroid and FlowDroid in 49 Genome Families}
\label{genomeSrcSink}
\end{figure*}

{\noindent \bf Evaluation on Genome Malware Apps:} Genome app set \cite{dissectingAndroid} consists of 49 malware families (with total of 1259 samples) where all members of a family exhibit similar malicious behaviors (with minor variations). Both Dexteroid and FlowDroid are applied on 49 families and their detection results are summarized in Figure \ref{genomeSrcSink}. The x-axis shows the top 20 sources which are leaked while y-axis shows the number of families which can leak that specific source data. The family here indicates that at least one of its members leaks the specific information. Dexteroid reports that the deviceID and location coordinates are leaked by 27 and 21 apps, respectively.

\begin{table*}[ht] 
\caption{Experimental Evaluation of Dexteroid and FlowDroid on 49 Genome Samples}
\centering 
\scriptsize
\begin{tabular}{l | l | l  l }
\hline 
& Item & DT & FD \\ 
\hline 
\parbox[t]{0.1mm}{\multirow{4}{*}{\rotatebox[origin=c]{90}{Setup}}} & \#Test Apps & 49 & 49 \\   
& $m$-way Permutation & $1$ & N/A\\   
&Max Time (min) &  10 & 10 \\   
&Source \& Sink API Set &  DT & DT \\   
&Warning Type & I & I  \\
\hline
\parbox[t]{2mm}{\multirow{8}{*}{\rotatebox[origin=c]{90}{Results}}} &\#Reported Warnings &  111 &  66 \\   
&\#True Warnings &  94 & 49 \\   
&\#Combined Warnings ($W_{CO}$) & 101 & 101\\   
&Recall(\%) & 93.1 & 48.5 \\   
&Precision(\%) & 84.7 & 74.2 \\  
&F-1 Score & 0.87 & 0.59  \\  
&\#Killed(\%) & 4 & 4 \\   
&\#Finished(\%) & 96 & 96 \\ 
\hline 
\multicolumn{4}{l}{
  \begin{minipage}{6cm}
    \scriptsize DT=Dexteroid, FD=FlowDroid,  I=Information leakage
  \end{minipage}
}\\
\end{tabular} 
\label{expResultsGenome} 
\end{table*} 

For 1259 Genome samples, Dexteroid produces 4352 warnings while FlowDroid produces 2621 warnings. To reduce the number of warnings to a reasonably verification level, we perform sampling at malware family level because of two reasons: (1) all members of a family exhibit similar malicious behaviors (with minor variations). (2) it removes biasness of a tool towards a specific family in the evaluation results. For example, AnserverBot has 187 samples while GGTracker has only 1 sample. Sampling at family level ensures that all samples are unique in nature and calculations are not affected by the family size. One sample is randomly picked from each family and 49 samples are collected. Table \ref{expResultsGenome} shows the experimental setup and the evaluation results for both tools. Dexteroid achieves high recall (93.1\%) and high precision (84.7\%) compared with FlowDroid's recall (48.5\%) and precision (74.2\%). Most of the apps for both tools are analyzed completely within 10 minutes. Dexteroid does not report any additional attack with $2$-way permutation or which starts from any supplementary callback.

{\noindent \emph{SMS And Phone-Number Leaking Malware}:} The 49 Genome families leak various kind of information (as shown in Figure \ref{genomeSrcSink}). However, many families are reported to leak highly-sensitive data such as user phone number and SMS messages \cite{dissectingAndroid}. Table \ref{smsPhoneAttacks} shows a list of malware families in which at least one member of the family leaks the specific information. Dexteroid detects reported SMS-leakage attacks for most of the families (9 out of 13), in addition to additional attacks for four other families. It misses detection for four families which use intent-communication (CoinPirate, NickyBot) or read data from SMS-content providers (Gone60, Nickyspy). Dexteroid's default handlers do not resolve input parameters for content provider APIs to find out the database name pointed by a Uri and hence lead to the missed attacks. Moreover, one false positive is produced for BeanBot malware. Similarly, Dexteroid detects phone number leaks for 10 out of 15 families (with no false positive) and further reports new (previously-unreported) leaks for 4 families (AnserverBot, BaseBridge, DroidKungFu4, RogueLemon). Five leaks are missed because of intent-communication or incomplete prototype implementation of components such as adapter. FlowDroid reports 5 malware families leaking SMS and 8 malware families leaking phone number as shown in Figure \ref{genomeSrcSink}.

\begin{table}[ht] 
\scriptsize
\caption{Detection of SMS And Phone-Number Leaking Families by Dexteroid in Genome Malware}
\centering 
\begin{tabular}{p{2cm} | p{1.2cm} p{1.2cm} p{1.2cm}  p{1.2cm} } 
\hline
{\multirow{2}{*}{Malware}} & \multicolumn{2}{c}{Sms Leakage} & \multicolumn{2}{c}{Phone No Leakage}  \\ \cline{2-5}  
  & Reported & Detected   &  Reported &  Detected \\ 
\hline 
ADRD &  - & \checkmark& -  & - \\  
AnserverBot & - & \checkmark &   -& \checkmark  \\  
BaseBridge &  - & \checkmark  &   -& \checkmark  \\  
BeanBot  & -& -& \checkmark   & \checkmark  \\        
BgServ  & - & - & \checkmark   & \checkmark\\    
CoinPirate  & \checkmark &  -& - & - \\  
Crusewin  & \checkmark & \checkmark & - & - \\   
DroidKungFu1 & - & - & \checkmark & \checkmark  \\   
DroidKungFu2 & - & - & \checkmark  &  -\\   
DroidKungFu3 & - & -& \checkmark  & \checkmark  \\   
DroidKungFu4 & - & - &  -& \checkmark  \\   
DroidKungFu5 & - & - & \checkmark  &  -\\  
Endofday  & - & - & \checkmark  & \checkmark  \\  
GamblerSMS & \checkmark  & \checkmark  &  -&  -\\
Geinimi & \checkmark &  \checkmark   & \checkmark  &  -\\ 
GGTracker & \checkmark  & \checkmark   &  \checkmark  & \checkmark  \\  
GingerMaster  & -  & - &  \checkmark  & \checkmark  \\  
GoldDream  & \checkmark  & \checkmark  &  \checkmark & \checkmark \\  
Gone60  & \checkmark  &  -& -  & - \\  
jSMSHider & - & - & \checkmark  &  -\\
NickyBot  & \checkmark  &  -& -& - \\  
Nickyspy  & \checkmark  &  -& -& - \\  
Pjapps  &   - & \checkmark &  \checkmark & \checkmark \\  
RogueLemon & \checkmark  & \checkmark   &   -& \checkmark \\  
SMSReplicator  & \checkmark  & \checkmark  & - & - \\  
Spitmo  & \checkmark   & \checkmark &   \checkmark  & \checkmark \\  
YZHC  &  -& - &  \checkmark  &  -\\
Zitmo & \checkmark  & \checkmark  & - & - \\  
\hline 
\end{tabular} 
\label{smsPhoneAttacks} 
\end{table} 
\subsubsection{Evaluation on DroidBench} \label{evalDroidBench}

DroidBench \cite{FlowDroid} is a suite of Android applications developed to evaluate correctness and completeness of an analysis tool such as Dexteroid. Most of the apps leak sensitive information such as deviceID. The apps evaluate a tool for both accuracy and precision against different challenges such as object and field sensitivity, life cycle callbacks, anonymous classes, AUI and miscellaneous callbacks, loops, inactive activity, and storing data into arrays and lists. We use latest version (2.0) of DroidBench apps. Since Dexteroid does not analyze intent-communication, Java reflection or implicit flows (see Section \ref{secMalBehavior}), applications that leak information using these features are removed from the evaluation set. Many benchmark apps contain flow to logging APIs (e.g., {\ttfamily Log.e()}), we add such APIs to our sink API set for the evaluation. Table \ref{droidBench} shows the evaluation results when both tools are run on the remaining 90 apps. Dexteroid achieves higher accuracy and precision as compared to FlowDroid. It accurately detects all the flows related to life cycle models of different components such as event (e.g., life cycle callback) ordering and multiple calls to {\ttfamily onStartCommand()} of the service component. It misses 11 leaks because of incomplete modeling of functionalities such as {\ttfamily Serialization}, {\ttfamily Parcel}, and {\ttfamily SharedPreferenceChanged}. These flows can be captured by implementing specific handlers for such APIs in the Dexteroid framework. Also, Dexteroid produces five false warnings: four for marking the whole collection objects (e.g., array, list and hashmap) as tainted (even if a value at a specific index is tainted only) and one for an invalid control flow in the exception handling. In comparison, FlowDroid misses flows because of incomplete handling of features such as static initializations, saved activity state in {\ttfamily onSaveInstanceState()} and {\ttfamily Arrays.toString()} API, in addition to the flows missed by the Dexteroid. Similarly, FlowDroid produces 8 false positives because of collection objects, invalid flows in exception handling, unregistered callbacks, and incorrect resolving of aliases in {\ttfamily Merge1} app.

\begin{table}[ht] 
\caption{Evaluation of Dexteroid and FlowDroid on DroidBench Apps}
\centering 
\scriptsize
\begin{tabular}{ l | l  l l  l l } 
\hline 
Tool & Total Flows & False Negatives & False Positives & Accuracy & Precision\\  
\hline
Dexteroid & 79 & 11 & 5 & 83.7\% & 93.1\% \\
FlowDroid & 79 & 22 & 8 & 69.3\% & 87.6\% \\
\hline 
\end{tabular} 
\label{droidBench} 
\end{table}

{\noindent \bf Additional Test Cases:} We develop six test apps based on event sequences derived from the reverse-engineered activity and service life cycle models and add these apps to the DroidBench suite. The apps evaluate if an analysis tool correctly models the component life cycle behaviors such as conditional flows between the callbacks, event-generated callbacks and different orderings of the event sequences. The evaluation results for both tools are shown in Table \ref{testcases}. Dexteroid accurately detects all the privacy leaks while FlowDroid detects one of them, but misses five privacy leaks. Moreover, the {\ttfamily ActivityEveSeq3} app in Table \ref{testcases} contains an attack given in the above motivating example. Dexteroid detects this attacks with $2$-way permutation but FlowDroid does not detect this attack.

\begin{table}[ht] 
\caption{Evaluation of Dexteroid and FlowDroid on Six Additional Test Case Apps}
\centering 
\scriptsize

\begin{tabular}{l P{1.7cm} P{6.7cm}  P{.7cm} P{.7cm} } 
\hline 
\# &App Name  & Event Sequence For The Attack &  DT &  FD\\ 
\hline
1 & ActivityEveSeq1 & createActivity               & \checkmark & $\times$ \\
2 & ActivityEveSeq2 & createActivity $\rightarrow$ hideActivityPartially $\rightarrow$ savStop $\rightarrow$ savRestart $\rightarrow$ savStop & \checkmark &  $\times$\\
3 & ActivityEveSeq3 & createActivity $\rightarrow$ hideActivityPartially $\rightarrow$ gotoActivity $\rightarrow$ overlapActivity $\rightarrow$restartActivity $\rightarrow$ confPR & \checkmark & $\times$\\
4 & ServiceEveSeq1 & createAndStart $\rightarrow$ bind $\rightarrow$ start & \checkmark & $\times$ \\
5 & ServiceEveSeq2 & createAndStart $\rightarrow$ bind $\rightarrow$  unbind $\rightarrow$ bind & \checkmark & $\times$ \\
6 & ServiceEveSeq3 & createAndBind $\rightarrow$ unbindAndDestroy & \checkmark & \checkmark\\
\hline 
\multicolumn{5}{l}{
  \begin{minipage}{8cm}
    \scriptsize DT=Dexteroid, FD=FlowDroid, \checkmark=detects, $\times$= does not detect
  \end{minipage}
}\\
\end{tabular} 
\label{testcases} 
\end{table} 

\subsubsection{Evaluation on Additional Callback Flows from Other Tools}  \label{EdgeMinerGator}

Experiments in Sections \ref{evalReadWorld}-\ref{evalDroidBench} show that Dexteroid can detect information leakages with high precision and high recall as compared to the existing tools such as FlowDroid. It considers life cycle callback sequences, UI callbacks and miscellaneous callbacks for the analysis. However, the manually-compiled list of UI and miscellaneous callbacks is potentially incomplete and may lead to many missed detections of malicious behaviors. For example, a call to {\ttfamily sort()} method (a registration method) of {\ttfamily Collections} class causes Android framework to implicitly invoke {\ttfamily compare()} method (a callback method) implemented by its {\ttfamily Comparator} class. The Android framework implicitly transfers program control flow from a registration method to its associated callback method. However, Dexteroid has only a limited mapping of such implicit control flow transitions (ICFT) (e.g., from {\ttfamily sort()} to {\ttfamily compare()} method) and may miss many detections during the analysis. Similarly, Dexteroid currently does not analyze callback sequences invoked by opening or dismissing dialog windows or menu items. In this section, we incorporate implicit control-flow transitions from EdgeMiner \cite{EdgeMiner} and callback sequences from Gator \cite{GATORWTG} into Dexteroid to evaluate their impact on Dexteroid's analysis results.

{\noindent \bf Integrating ICFTs from EdgeMiner:} EdgeMiner \cite{EdgeMiner} statically analyzes Android framework to automatically identify a complete set of implicit control flow transitions (ICFTs) which can occur in Android application space (and framework space). Each ICFT consists of a callback method and its associated registration method. An ICFT can be of two types: (1) synchronous, and (2) asynchronous. In a synchronous ICFT, a callback method (e.g., {\ttfamily compare()} method) is invoked synchronously as soon as its registration method (e.g., {\ttfamily sort()} method) is invoked while in an asynchronous ICFT, the callback method (e.g., {\ttfamily onClick()} method) is invoked after some delay to the invocation of its registration method (e.g., {\ttfamily setOnClickListener} method). EdgeMiner's analysis, however, does not aim to identify life cycle callbacks of Android components.
	
	We apply EdgeMiner tool, thanks to its open-source nature, on Android 4.4 to generate 5,655,548 ICFTs (i.e., registration and callback method pairs). We further use EdgeMiner to separate synchronous and asynchronous ICFTs and modify Dexteroid to incorporate callbacks of these ICFTs. Upon identifying a registration method of a synchronous ICFT during analysis, Dexteroid analyze its callback method, if present in the program, using Algorithm \ref{methodAnalyzer} (see Section \ref{taintAnalysisSec}). For asynchronous ICFTs, it considers their callbacks as permutation units and analyzes them separately after the callback sequence of createActivity event. This is because all such asynchronous callbacks can be invoked randomly when activity is in a StaticPostResumed state. It should be noted that in both cases, Dexteroid matches method-signatures to identify registrations and callbacks for the analysis, instead of just relying upon the method name which may lead to imprecise analysis results.
	
	We perform two experiments with Dexteroid after integrating ICFTs from EdgeMiner. The first experiment aims to confirm that after integration, Dexteroid can detect attacks launched by synchronous and asynchronous ICFTs. Due to unavailability of six sample applications used by EdgeMiner \cite{YCao}, we develop four testbed applications that leak sensitive information using synchronous and asynchronous ICFTs. Dexteroid successfully detects all privacy leaks in these testbed applications. 
	
	In the second experiment, we further evaluate impact of EdgeMiner ICFTs on analysis results by running Dexteroid on 158 apps of experiment\#2 in Table \ref{expResults}. Each app is analyzed with 1-way permutation for a maximum of 30 minutes. Dexteroid reports 9 new warnings by 6 apps, in addition to the 155 warnings reported earlier in experiment\#2 of Table \ref{expResults}. The new warnings leak information such as deviceID, geo-location coordinates, and network operator name using callbacks such as {\ttfamily onKeyUp()}, {\ttfamily onViewCreated()}, {\ttfamily onExit()}, and {\ttfamily onAttachedToWindow()}. After manual verification, we find seven of these warnings are true positives. Moreover, all apps are analyzed completely within the time limit of 30 minutes.

{\noindent \bf Handling Callback Sequences from Gator:} Gator \cite{GATORWTG} performs static analysis to build a window transition graph (WTG) of an Android application. The nodes in WTG represent windows (e.g., activity, dialog) and edges represent transitions between the windows, triggered by callbacks executed in the main UI thread. It accurately models Android's "back stack" \cite{BackStack} to determine possible valid transitions among the windows. Using widget events, and five default events (back, rotate, power, home, menu), it determines callback sequences which are invoked by Android for windows, and their transitions. Since the WTG is aimed to be primarily used for GUI-testing, it generates callback sequences only for GUI-elements (e.g., activity, dialog and menu). It does not derive callback sequences for other components such as service, broadcast receiver and asynchronous task.

To perform experiments with Dexteroid, we use authors' \cite{ARountev} recommended implementation of Gator (version 3.0) to obtain callback sequences from the WTG because it handles more events and more of the Android semantics as compared to its predecessor version 2.0 \cite{GatorICSE} (see Section \ref{relatedWork}). Furthermore, we implement a parser to extract callback sequences for individual components from the WTG output because due to inter-component communication analysis, the produced callback sequences for a component (e.g., activity) contain callbacks from other components. With Gator callback sequences, we perform two types of experiments: (1) by replacing Dexteroid callback sequences with Gator's callback sequences, and (2) by integrating Gator callback sequences into Dexteroid callback sequences.

For the first type of experiment (i.e., with Gator callback sequences alone), Dexteroid is first run with 2-way permutation on six additional test case apps discussed in Section \ref{evalDroidBench}. It reports warning for only one app (ActivityEveSeq1) which uses one callback sequence to launch the attack. We then run Dexteroid with 1-way permutation on 158 apps of experiment\#2 in Table \ref{expResults}. The experiment is run with 30 minutes each per application so that all apps are analyzed completely within the time limit. Dexteroid reports a total of 120 warnings which all are reported earlier in 155 warnings of experiment\#2 of Table \ref{expResults}. This reduction in warnings is because (1) Gator's callback sequences are specific to GUI-elements only (e.g., activity, dialog and menu). Out of 155 warnings, a total of 27 warnings were earlier contributed by service and broadcast receiver components. Gator does not produce callback sequences for these components. (2) For the activity component, Dexteroid now reports 120 warnings (as compared to earlier-reported 128 warnings) because of reduced number of Gator callback sequences for the analysis.
	
For the second type of experiment (i.e., with Gator callback sequences integrated into Dexteroid), we repeat the above experiment on 158 apps. Dexteroid reports the same 155 warnings without any new warning. This is partly because most of the events used by Gator to derive callback sequences for activities are already covered by the Dexteroid. For Dexteroid, the new callback sequences for analysis come only from dialogs, option menus and context menus for which it does not report any new warning. All apps are completely analyzed under 30 minutes in the above experiments.

\subsection{Detection of SMS-Sending Attacks} 

An app can send SMS to premium-rate numbers which may cost users a lot in their monthly bills. Dexteroid reports warnings for apps which can send SMS to hard-coded numbers defined in the app code, or send automatic reply to an incoming SMS message. FlowDroid does not aim to detect such attacks. Thus, we present our evaluation with Dexteroid in the following sections.

{\noindent \bf Evaluation on Google Play Apps:} Dexteroid analyzes 1526 GPlay apps with $1$-way permutation and 10 minutes per app configuration to detect SMS-sending attacks. During the analysis, it reports a warning if a hard-coded number or SMS sender's phone number obtained from an API such as {\ttfamily SmsMessage.getOriginatingAddress} API is passed to the recipient parameter of an SMS sending API (e.g., first parameter of {\ttfamily sendTextMessage()} API). Dexteroid reports that two Google Play apps ({\ttfamily com.metropcs.service.vvm} and {\ttfamily com.saavn.android}) can send SMS to hard-coded numbers. We manually verified that both warnings are part of legitimate functionality of the apps. Furthermore, Dexteroid successfully detects that three apps ({\ttfamily com.blendr.mobile}, {\ttfamily com.badoo.mobile}, and {\ttfamily com.hotornot.app}) can send automatic reply to an incoming SMS in {\ttfamily onReceive()} method of a broadcast receiver. These apps contain the same in-app billing library which contains this functionality.

{\noindent \bf Evaluation on Genome Malware Apps:} Many of the Genome malware families are reported to launch SMS-sending attacks \cite{dissectingAndroid}. Table \ref{smsAttacks} shows analysis results of Dexteroid on Genome malware. It successfully detects SMS-sending attacks to hard-coded numbers in FakePlayer, HippoSMS, and Zsone malware, and SMS-reply to incoming messages in Endofday, GPSSMSSpy, and GGTracker malware. Moreover, Dexteroid finds that Pjapps, CoinPirate, Nickyspy, Geinimi and DogWars malware send SMS to hard-coded numbers but these behaviors were not reported in \cite{dissectingAndroid}. However, Dexteroid detects only those attacks which obtain their recipient numbers from within the app code. For example, it does not detect attacks in Jifake and Crusewin malware because their hard-coded numbers are given in configuration files. Similarly, SMSReplicator and Walkinwat obtain hard-coded numbers from a {\ttfamily cursor} object. Other malware such as NickyBot, Nickyspy, Pjapps, and YZHC malware communicate with command and control (C\&C) server to obtain recipient numbers. Such attacks are harder to detect by a static analysis tool because the required information is not available in the app code.

\begin{table}[ht] 
\scriptsize
\caption{Detection of SMS-Sending Attacks by Dexteroid in Genome Malware}
\centering 
\begin{tabular}{p{2cm} | p{.9cm} p{.9cm} p{.9cm} p{.9cm} } 
\hline
{\multirow{2}{*}{Malware}} & \multicolumn{2}{l}{Sms to Hard-coded } & \multicolumn{2}{c}{Sms Reply to}  \\  
        &  \multicolumn{2}{c}{Numbers}    &  \multicolumn{2}{c}{Incoming Messages}\\ \cline{2-5} 
  & Reported & Detected   &  Reported &  Detected \\
\hline
CoinPirate  & - & \checkmark & - & - \\  
Crusewin  & - & - & \checkmark & - \\  
DogWars  & - & \checkmark & \checkmark  & - \\  
Endofday  & - & - & \checkmark  & \checkmark  \\  
FakePlayer  & \checkmark  & \checkmark & - & -\\  
Geinimi & - & \checkmark  & - & - \\
GGTracker & \checkmark  & - & - & \checkmark  \\  
GPSSMSSpy  & - &- & \checkmark  & \checkmark \\  
HippoSMS  & \checkmark  & \checkmark & - & - \\  
Jifake  & \checkmark  & - & - & - \\  
NickyBot  & - & -& \checkmark  &  -\\  
Nickyspy  & - & \checkmark & \checkmark  & - \\  
Pjapps  & - & \checkmark & - & \checkmark \\  
RogueSPPush  & \checkmark  & - & - & \checkmark\\  
SMSReplicator  & - & - & \checkmark  & - \\  
Walkinwat  & - & - & \checkmark  & -\\  
YZHC  & \checkmark  & - & - & - \\  
Zsone  & \checkmark  & \checkmark & - & -\\  
\hline
\end{tabular} 
\label{smsAttacks} 
\end{table} 

\subsection{Performance Evaluation} \label{performanceEvaluation}

All of our experiments are run on a desktop machine running Ubuntu 14.04 with AMD Phenom II quad core processor, and 16 GB of memory. We run both tools on 1526 GPlay apps with maximum allowed time of 10 minutes per app. Dexteroid is not able to complete its analysis for 30\% of the apps (470 apps) while FlowDroid's analysis is killed for 42\% of the apps (640 apps). The killed apps by both tools can be different and can adversely affect the performance evaluations. To make an accurate comparison, we choose only those applications for which both tools complete their analyses without running out of time or memory. We use 158 GPlay apps from Table \ref{expResults} and 47 Genome apps from Table \ref{expResultsGenome}. Table \ref{perfrmEvaluation} shows the experimental setup and evaluation results. It shows total CPU time in minutes for each experiment and average CPU time per app in seconds taken by each tool for the analysis. We measure CPU time in place of wall-time because CPU time measures the amount of time for which CPU has been used to process all instructions of a program. The CPU time of a multi-thread program (such as FlowDroid) can be significantly greater than its wall-time. For example, for 158 GPlay apps shown in Table \ref{perfrmEvaluation}, FlowDroid's CPU time is 195.49 minutes while its wall-time is only 86.8 minutes. In comparison, Dexteroid's CPU time for 158 apps is 105.5 minutes. For both sets of apps, Dexteroid takes less time for analysis as compared to FlowDroid. However, we expect Dexteroid to take more time for higher order $m$-way permutation analysis.

\begin{table}[ht] 
\caption{Performance Evaluation of Dexteroid and FlowDroid}
\centering 
\scriptsize
\begin{tabular}{ l | l | p{.85cm} p{.85cm} | p{.85cm}  p{.85cm}  }
\hline
& {\multirow{2}{*}{Item}} & \multicolumn{2}{c|}{Experiment\#1}  & \multicolumn{2}{c}{Experiment\#2}  \\ \cline{3-6}
   & &  DT & FD & DT & FD \\  
\hline
\parbox[t]{0.1mm}{\multirow{2}{*}{\rotatebox[origin=c]{90}{Setup}}} & \#Test Apps & 158\dag & 158\dag  & 47$\star$ & 47$\star$ \\ 
& $m$-value & 1 & N/A &  1 & N/A    \\  
& Max Analysis Time/App (min) &  10 & 10 & 10 & 10  \\ 
\hline
\parbox[t]{2mm}{\multirow{3}{*}{\rotatebox[origin=c]{90}{Results}}} &Total CPU Time (min) & 105.5 & 195.49 & 20.08 & 38.69 \\
&Avg. CPU Time/App (sec) & 40.06 & 75.23 & 25.63 & 49.39 \\
&\#Killed(\%) & 0 & 0  & 0 & 0 \\  
&\#Finished(\%) & 100 & 100 & 100 & 100 \\ 
\hline
\multicolumn{6}{l}{
  \begin{minipage}{6cm}
    \scriptsize FD=FlowDroid, DT=Dexteroid, \dag =GPlay, $\star$=Genome
  \end{minipage}
}\\
\end{tabular} 
\label{perfrmEvaluation} 
\end{table}

\section{Related Work} \label{relatedWork}

There is a large body of work on Android security including static analysis \cite{LeakMiner, FlowDroid, AppoScopy, myDroid, Chex, DroidSafe, AndroidLeaks, Elish2015255, ScanDal, AAPL, EdgeMiner, Anadroid, ICCTAFD, cmuICCTA, AppIntent} and dynamic analysis \cite{TaintDroid, VetDroid, DroidScope, DroidPF, DroidBox} but we discuss only closely-related work. 

FlowDroid \cite{FlowDroid} and DroidSafe \cite{DroidSafe} are static taint analysis tools to detect privacy leakages in Android apps, and have been discussed in detail in Section \ref{secComparison}. In addition, DroidSafe analyzes inter-component communication also. However, given the reverse-engineered life cycle models, we need more careful approach to model and analyze inter-component communication which we leave as part of our future work. LeakMiner \cite{LeakMiner} builds a dummy main method (like FlowDroid \cite{FlowDroid}) and uses static taint analysis to detect information leaks. It includes supplementary callbacks in the CFG but it does not consider the guard conditions for the accurate analysis. Moreover, its context-insensitive approach can produce many false alarms. AndroidLeaks \cite{AndroidLeaks} performs static taint analysis based on System Dependence Graph (SDG) of WALA \cite{WALA}. It uses context-insensitive overlaying for heap dependencies in the SDG which leads to object-insensitive analysis. CHEX \cite{Chex} defines app \emph{split} as the app code reachable from an \emph{entry point} (e.g., life cycle callback). It performs data flow analysis on permutation of app \emph{splits} to detect component hijacking vulnerabilities, which can be extended to detect information leakage in Android apps. As expected, this app \emph{split} permutation may produce false positives because of the infeasible sequences \cite{Chex}. ScanDal \cite{ScanDal} is an abstract interpretation framework which works on Dalvik bytecode of the apps to find privacy leaks. It performs context-sensitive (with context depth=1) and hybrid of flow-sensitive and flow-insensitive analysis. In \cite{Elish2015255}, authors profile statically-extracted \emph{user-trigger dependence} to detect Android malware. AAPL \cite{AAPL} uses enhanced data flow analysis techniques to find privacy disclosures and then uses peer-voting mechanism to validate warnings reported for an application. 

Other techniques such as \cite{myDroid, ICCTAFD, cmuICCTA, AppIntent, ICCDamien} analyze inter-component communication for the analysis. ScanDroid \cite{myDroid} finds inter-component and inter-app data flows in the application. Based on WALA \cite{WALA}, however, it requires source code or JVML bytecode of the applications to perform analysis. IccTA \cite{ICCTAFD} and DidFail \cite{cmuICCTA} combine FlowDroid \cite{FlowDroid} and Epicc \cite{Epicc} to detect inter-component privacy leaks. Most of these techniques rely on Android-supplied life cycle models which lack possible flows between the callbacks and the guard conditions to prevent infeasible flows. We plan to extend the above reverse-engineered life cycle models to handle inter-component communication for analysis in our future work.

Dynamic analysis tools perform analysis by executing apps either in an instrumented Android OS \cite{TaintDroid, VetDroid} or in a virtualization-based environment \cite{DroidScope, DroidPF, DroidBox}. TaintDroid \cite{TaintDroid} performs taint flow analysis at multiple levels in Android OS and apps to detect potential information leakage in Android apps. DroidScope \cite{DroidScope} provides APIs to facilitate custom analysis at different levels: hardware, OS, and application level. DroidPF \cite{DroidPF} applies model checking techniques in a mocked-up Android OS to verify security and privacy properties of Android apps. Andromaly \cite{Andromaly} employs host-based anomaly detection to detect Android malware. 

In the realm of application life cycle models, Franke et al \cite{Franke2011, serviceLCMFrank} use same technique as discussed above to perform reverse-engineering on Android, iOS and Java ME applications. Their activity life cycle model developed for Android 2.2---a pre-Honeycomb version---contains some flows between the callbacks which do not occur in the Android's post-Honeycomb versions. AndroLift \cite{AndroLift} helps developers to monitor, implement and test life cycle related properties of Android applications. While Dexteroid relies on input events and dynamic monitoring to discover callback flows, Gator \cite{GatorICSE} performs context-sensitive static analysis to determine control-flow between callbacks of an Android app. It builds a callback control-flow graph (CCFG) which precisely captures both intra-component and inter-component callback flows in the code. However, Gator is specific to only user-driven components such as activities, dialogs, and menus. It does not handle other components such as service and broadcast receiver. Furthermore, from an activity life cycle, Gator considers only two callbacks ({\ttfamily onCreate()} and {\ttfamily onDestroy()}) in the CCFG, and identifies other flows (e.g., {\ttfamily onClick()}) between these two callbacks. Authors' most recent version of Gator implementation \cite{GATORWTG}, however, includes more life cycle callbacks and uses many widget and default events to produce window transition graph (WTG). This WTG can be used to derive callback sequences for security analysis purposes (see Section \ref{EdgeMinerGator}).

 \section{Conclusion and Future Work}

In this paper, we demonstrated that attacks can be designed to bypass early approaches, which are developed based on Android-supplied life cycle models. We perform reverse-engineering to reconstruct activity and service life cycle models and systematically derive event sequences from these models. In addition, the high-level event sequences offer great value in selecting callback sequences for the taint analysis. A series of experiments performed on Google Play apps and known Genome Malware apps show better performance in terms of precision and recall as compared with previous static analysis tool. Further experiments on DroidBench apps and six additional test case apps validate the effectiveness of the approach.

Dexteroid also has some limitations. First, as it does not handle inter-component or inter-app communication. Given the above reverse-engineered life cycle models, a careful modeling approach is required to handle inter-component communication for the analysis which we leave for our future work. Second, Dexteroid currently does not detect attacks based on implicit flows, Java reflection and native code. We would like to improve Dexteroid to detect such attacks in the future. Third, our current model detects only two types of attacks: (1) information leakage, and (2) sending SMS to premium-rate numbers. In the future, we are interested in detecting other attacks such as leaking a recorded voice call or making unintended phone calls. 

 \section*{Acknowledgements}

We thank the two anonymous reviewers whose constructive and valuable feedback helped improve quality of our paper. We appreciate Yinzhi Cao and Atanas Rountev for providing us great help in using EdgeMiner and Gator tools, respectively. We also thank Xuxian Jiang for sharing Android malware dataset with us for our experiments.

\renewcommand{\thesection}{}
\makeatletter
\def\@seccntformat#1{\csname #1ignore\expandafter\endcsname\csname the#1\endcsname\quad}
\let\sectionignore\@gobbletwo
\let\latex@numberline\numberline
\def\numberline#1{\if\relax#1\relax\else\latex@numberline{#1}\fi}
\makeatother


  \bibliographystyle{elsarticle-num} 

  \bibliography{dexteroid}

\end{document}